\documentclass[usenatbib]{mnras}

\usepackage{newtxtext}

\usepackage[T1]{fontenc}
\usepackage{ae,aecompl}

\usepackage{graphicx}	
\usepackage{amsmath}	
\usepackage{amssymb}	

\usepackage{ulem}
\usepackage{soul}

\newcommand{\code}[1]{\texttt{#1}}

\newcommand{\spr}{\mbox{$s$-process}}

\newcommand{\ipr}{\mbox{$i$-process}}
\newcommand{\iprn}{\mbox{$i$ process}}

\newcommand{\rpr}{\mbox{$r$-process}}
\newcommand{\rprn}{\mbox{$r$ process}}

\newcommand{\unitspace}{\ensuremath{\,}}
\newcommand{\usp}{\unitspace}

\newcommand{\unitstyle}[1]{\ensuremath{\mathrm{#1}}}
\newcommand{\power}[2]{\ensuremath{{#1}^{#2}}}
\newcommand{\ee}[1]{\ensuremath{\times 10^{#1}}}

\newcommand{\centi}{\unitstyle{c}}

\newcommand{\meter}{\unitstyle{m}}

\newcommand{\second}{\unitstyle{s}}

\newcommand{\Kelvin}{\unitstyle{K}}
\newcommand{\K}{\Kelvin}  

\newcommand{\cm}{\centi\meter}
\newcommand{\gram}{\unitstyle{g}}

\newcommand{\grampercc}{\gram\usp\power{\cm}{-3}} 

\newcommand{\erg}{\unitstyle{erg}} 

\newcommand{\ergspergs}{\erg\unitspace\power{\gram}{-1}\unitspace\power{\second}{-1}} 

\newcommand{\Msun}{\ensuremath{\unitstyle{M}_\odot}}
\newcommand{\Lsun}{\ensuremath{\unitstyle{L}_{\odot}}}
\newcommand{\Rsun}{\ensuremath{\unitstyle{R}_{\odot}}}

\newcommand{\yr}{\unitstyle{yr}}        

\input{./vectors}
\newcommand{\lFig}[1]{{\label{fig:#1}}}
\newcommand{\lEq}[1]{{\label{eq:#1}}}
\newcommand{\lTab}[1]{{\label{tab:#1}}}

\newcommand{\Tabff}[1]{{\ref{tab:#1}}}
\newcommand{\Tab}[1]{{Table~\Tabff{#1}}}

\newcommand{\pan}[1]{{\textit{#1}}}

\newcommand{\FIGFF}[2]{{\ref{fig:#2}\pan{#1}}}

\newcommand{\FIG}[2]{{Fig.~\FIGFF{#1}{#2}}}
\newcommand{\Fig}[1]{{\FIG{}{#1}}}

\newcommand{\Eqref}[1]{{\ref{eq:#1}}}
\newcommand{\Eqff}[1]{{(\Eqref{#1})}}

\newcommand{\Eq}[1]{{Eq.~\Eqff{#1}}}

\usepackage[usenames,dvipsnames]{color}

\newcommand{\Rshell}{\texttt{Rad-Shell}}
\newcommand{\Cshell}{\texttt{Conv-Shell}}
\newcommand{\Ccore}{\texttt{Conv-Core}}
\newcommand{\Rcore}{\texttt{Rad-Core}}

\title[ H-He in Pop III ]{Convective H-He Interactions in Massive Population III Stellar Evolution Models}

\author[O.~Clarkson et.~al]{
O.~Clarkson,$^{1,2}$\thanks{E-mail: oclark01@uvic.ca}
F.~Herwig,$^{1,2}$
\\
$^{1}$Department of Physics \& Astronomy, University of Victoria, P.O. Box 3055 Victoria, B.C., V8W 3P6, Canada\\
$^{2}$Joint Institute for Nuclear Astrophysics - Center for the Evolution of the Elements (JINA-CEE)\\
}

\date{Accepted XXX. Received YYY; in original form ZZZ}

\pubyear{2020}

\begin{document}

\label{firstpage}
\pagerange{\pageref{firstpage}--\pageref{lastpage}}
\maketitle
\begin{abstract}
In Pop III stellar models convection-induced mixing between H- and He-rich burning layers can induce a burst of nuclear energy and thereby substantially alter the subsequent evolution and nucleosynthesis in the first massive stars. We investigate H-He shell and core interactions in 26 stellar evolution simulations with masses $15 \-- 140\,\mathrm{M}_{\odot}$, using five sets of mixing assumptions. In 22 cases H-He interactions induce local nuclear energy release in the range $ \sim 10^{9} \-- 10^{13.5}\,\mathrm{L}_{\odot}$. The luminosities on the upper end of this range amount to a substantial fraction of the layer's internal energy over a convective advection timescale, indicating a dynamic stellar response that would violate 1D stellar evolution modelling assumptions. We distinguish four types of H-He interactions depending on the evolutionary phase and convective stability of the He-rich material. H-burning conditions during H-He interactions give $^{12}\mathrm{C}/^{13}\mathrm{C}$ ratios between $\approx 1.5$ to $\sim1000$ and [C/N] ratios from $\approx -2.3 $ to $\approx 3$ with a correlation that agrees well with observations of CEMP-no stars. We also explore Ca production from hot CNO breakout and find the simulations presented here likely cannot explain the observed Ca abundance in the most Ca-poor CEMP-no star. We describe the evolution leading to H-He interactions, which occur during or shortly after core-contraction phases. Three simulations without a H-He interaction are computed to Fe-core infall and a $140\,\mathrm{M}_{\odot}$ simulation becomes pair-unstable. We also discuss present modelling limitations and the need for 3D hydrodynamic models to fully understand these stellar evolutionary phases. 
\end{abstract}

\begin{keywords} 
stars: massive -- evolution -- Population III,  nuclear reactions, nucleosynthesis, abundances 
\end{keywords}


\section{introduction} \label{Section 1} 
Metal-poor and CEMP (Carbon-Enhanced Metal-Poor) \citep{2005ARA&A..43..531B} stars provide a wealth of information regarding the nature of early galactic chemical enrichment. It is often supposed that the most Fe-poor of these stars have been enriched by a single Population III (Pop III) star \citep{2015ARA&A..53..631F}. In order to replicate the abundances of some of the most Fe-poor stars it has been found that some degree of mixing between H- and He-burning layers during the stars lifetime is likely needed \citep[e.g.,][]{2003ApJ...594L.123L, 2015A&A...580A..32M}, as this mixing activates nucleosynthetic pathways otherwise closed. While such nucleosynthesis appears necessary to reproduce the abundance patterns in many Fe-poor stars, as of yet, there has been no systematic investigation into the impact of such events on a set of 1D Pop III stellar models.

1D stellar evolution simulations suggest that massive Pop III stars may have undergone mixing events between H- and He-burning layers. This was first described by \cite{1982ASIC...90.....R} for a very massive Pop III stellar model. \cite{2001A&A...371..152M} found H-He interactions\footnote{Throughout this paper we refer to all events where H- and He-rich material mix as H-He interactions rather than proton/H-ingestion or any other name regardless of the stellar site, unless comprehensibility demands it.} between H shells and He cores but dismissed their results as \textit{physically unsound} due to the lack of coupling of mixing and burning in their code at the time. 
 
\cite{2010ApJ...724..341H} describe such events occurring in a 25\usp\Msun\ model after core C depletion in a set of Pop III massive stellar models used to produce supernova yields. They observed the convective He-burning shell and radiative envelope to mix. This was reported to produce large amounts of N due to the boost in CNO at the base of the H envelope as He-burning ashes were made available. In an 80\usp\Msun\ model during core He burning, they found the He core encroached upon the H shell, ultimately mixing H downward to create N which then convected downward, producing $^{22}\mathrm{Ne}$. They considered these events and the nucleosynthetic by-products as uncertain yet unavoidable. 

In another set of supernova yields, \cite{2012ApJS..199...38L} note that their stellar models of massive, Pop III stars encounter interactions between H and He shells. They report such mixing for stellar models of 25, 30 and 35\usp\Msun\, again, typically increasing the abundance of primary N in the H shell in addition to the high [Na/Mg] reported in a previous work endeavouring to explain the most Fe-poor star at the time, HE 0107-5240 \citep{2003ApJ...594L.123L}. 

\cite{2018MNRAS.480..538R} encountered H-He interactions in simulations of $1\usp\Msun \-- 25\usp\Msun$  from $Z = 0.006$ and $ Z = 0.0001$ and report neutron densities from such events of $\log N_n \sim 11-12$ resulting from the activation of the $^{13}\mathrm{C}(\alpha, n)^{16}\mathrm{O}$ neutron source. 
 
 It has been suggested that nucleosynthesis patterns in some CEMP-no stars --- CEMP with no overabundance of $s$- or \rpr\ elements, may be the result of a late mixing process between H and He-layers, potentially resulting in the variety of $^{12}\mathrm{C}/^{13}\mathrm{C}$ ratios observed in CEMP-no stars \citep{2016A&A...593A..36C}. In a study on the effect of rotation on 1D Pop III models, \cite{2008A&A...489..685E} reported H- and He-mixing during core He burning for both rotating and non-rotating models but attributed it to rotational mixing where there was a higher frequency of the mixing event. 
 
In the pursuit of understanding the chemical abundances of some CEMP stars, the \iprn\, has been suggested as a possible explanation. The \iprn\, is a neutron-capture process with neutron densities, $N_\mathrm{n}$,  intermediate between those of the $s$ and \rprn es. This process was originally discussed by \cite{1977ApJ...212..149C} for low-mass stars undergoing He shell flashes and has been invoked to explain several abundance patterns more recently including the post-AGB star V4334 Sagittarii and CEMP-r/s stars \citep[e.g.][]{2011ApJ...727...89H, 2014nic..confE.145D,2019MNRAS.488.4258D, 2016ApJ...831..171H, 2019ApJ...887...11H}. The \iprn\ occurs in convective-reactive regimes---when mixing and nucleosynthesis timescales are comparable. For this reason, the Damh{\"o}hler number, or $\mathrm{Da}$ has been used to describe \ipr\ conditions  as it is the dimensionless ratio of the nuclear reaction timescale relative to the mixing timescale \citep{2011ApJ...727...89H}. \ipr\ conditions are characterized by $\mathrm{Da} \approx 1$.

 \cite{2016MNRAS.455.3848J} studied models of super-AGB stars of $\mathrm{Z} = 10^{-5} - 0.02$.  In these simulations the convective H-envelope is found to burn corrosively into core He-burning ashes after the second drege-up, similar to what has been seen in $9\Msun$ and $7\Msun$ models of \cite{1997ApJ...485..765G} and \cite{2012ApJ...757..132H}, respectively. After this corrosive burning phase a dredge-out event occurs \citep[see also][]{1999ApJ...515..381R, 2013A&A...557A.106G, 2014MNRAS.441..582D}. This happens where He burning is sufficiently strong to induce convection in the He shell during the second dredge-up, and the descending H envelope and expanding He shell merge. \cite{2016MNRAS.455.3848J} found these shell merger events, where protons descend into the He shell reacting with C via the $^{12}\mathrm{C}(p,\gamma)^{13}\mathrm{N}$ reaction, lead to luminosities in excess of $10^{9}\usp \Lsun$ and posit that such an event would lead to the \iprn\ followed by mass-ejection events.
 
 \cite{2018MNRAS.474L..37C, 2019MNRAS.488..222C} investigated a H-He interaction in a $45\usp\Msun$ Pop III stellar model, leading to \iprn\ conditions and compared nucleosynthesis results with the elemental abundances of the three most Fe-poor stars at the time, SMSS J031300.36-670839.3 \citep{keller:14}, HE1327-2326 \citep{2006ApJ...652.1585F}, and HE0107-5240 \citep{2004ApJ...603..708C}. With luminosities of $10^{13}\usp\Lsun$,  mixing length theory (MLT)  \citep{1958ZA.....46..108B} is unable to accurately predict the resulting nucleosynthesis and therefore, exploratory single-zone nucleosythesis calculations using temperatures and densities found in the He shell were performed. We found that the light-element signature---trends of high [Na/Mg], [Mg/Al] and [Al/Si]---are naturally reproduced by the \iprn. Here the odd-Z elements are made by n-captures as well as Ca. Similar trends were also recovered in Ti-Mn which has been seen in several other CEMP-no stars. 

Nucleosynthesis in an $80\usp\Msun$ model was investigated using the same methods and predict these events can produce elements heavier than Fe without violating observed upper limits in CEMP-no stars \citep{Clarkson2019}. Similar to \cite{2016MNRAS.455.3848J}, \cite{2018MNRAS.474L..37C} suggested that mass-ejection events might result from H-He interactions, not unlike the precursor events that lead to the circumstellar material observed in Type IIn SNe or other pre-supernova outburst events \citep[see][]{2011MNRAS.415..773S}. This scenario does not include a faint supernova, as Pop III stars with masses $\sim 40 \-- 100\usp\Msun$ are expected to collapse into a black hole directly with no supernova explosion \citep{2003ApJ...591..288H}. More recent studies including the neutrino-driven explosion mechanism paint a more complicated picture regarding the fate of massive stars of zero intial metallicity \citep[e.g.][]{2020ApJ...888...91E}.  \cite{2020MNRAS.492.2578S} suggest that the upper mass limit for a successful core collapse supernova explosion may be closer to $\sim 20\,\Msun$, which seems to scale only weakly with metallicity \citep{2014ApJ...783...10S}, with lower metallicity stars producing more black holes, generally. The ultimate fate of these stars is currently an unsolved problem, requiring 3D hydrodynamic simulations including a number of physical ingredients \citep{2017RSPTA.37560271C,2017MNRAS.472..491M}.

\cite{2018ApJ...865..120B} also looked at mixing of H with He shells of massive Pop III and low-Z stellar models. During post-processing, to reproduce heavy-element abundances of CEMP-no stars, they induce H-He interactions by injecting protons to the top of the He shell. In doing so, they are able to reproduce the overall heavy element abundances in CEMP-no stars. 

In low and intermediate-mass stellar models, interactions between H and He layers have been studied by a multitude of authors and are well documented in the literature. These events have been found to occur in several stellar sites for different metallicities, masses and evolutionary phases. The He-shell flash \citep[e.g.][]{1990ApJ...349..580F, 2009PASA...26..139C,2010MNRAS.405..177S},  hot dredge-up \citep{2004ApJ...605..425H,2004A&A...421L..25G}, and thermal pulses and interpulse periods \citep{1989ApJ...340..966H, 2012ApJ...747....2L} in Pop III and low-Z AGB stars, the He core flash \citep{2010A&A...522L...6C,2001ApJ...559.1082S}, and rapidly-accreting white dwarfs \citep{2019MNRAS.488.4258D} are all sites where H-He interactions have been found to occur.

Studies involving the impact of mixing assumptions in H-He interactions have been done for the case of low and intermediate mass stars. For these stars, early studies focused on the effects of mixing rate of protons into a He-rich environment and splitting of the He-rich convection zone \cite[][and references therein]{1986MNRAS.223..683M}.
More recently, the focus has moved to the influence of convective boundary mixing (CBM). \cite{2006A&A...449..313M} explored the influence of convective mixing theories and CBM in post-AGB models, finding that born-again episodes were altered by both considerations. Studies have also found that both CBM and mass can effect the depth of dredge up episodes \citep{2004ApJ...605..425H, 2009MNRAS.396.1046L}. \cite{2016MNRAS.455.3848J} studied CBM efficiency for the envelope and pulse driven convection zone in SAGB stars. They found that for all values of CBM efficiency H-He interactions were encountered frequently, with higher values generally leading to higher H-burning luminosities.

Convective shell interactions (CSI) between other burning shells during advanced burning stages in massive stars can lead to unique nucleosynthesis and asymmetries in 3D hydrodynamic simulations possibly aiding supernova explosions \citep{2019arXiv190504378Y,2020MNRAS.491..972A}. 

In this paper, we explore the impact of a range of commonly adopted convective boundary mixing assumptions on Pop III massive star evolution models. Thereby this paper complements the recently emerged effort of the community to better understand the impact of macro- and microphysics uncertainties on model predictions. For example, \cite{2020MNRAS.tmp.1745K} investigated the impact of
  convective boundary mixing parameters main-sequence (MS) and He-burning evolution for three different stellar mass massive star models at solar metallicity. \cite{2018ApJS..234...19F} have investigated the impact of nuclear reaction rate uncertainties for a $15\usp\Msun$ stellar evolution model. \cite{2014ApJ...783...10S} included separate sensitivity studies on the impact of resolution and mixing, in the form of semiconvection and CBM, on the compactness parameter \citep{2011ApJ...730...70O} in massive pre-supernova stellar models. In a similar vein, \cite{2019MNRAS.484.3921D} examined how a varying CBM parameter affects the late-time evolution of a $25\usp\Msun$ stellar model. \cite{2016ApJS..227...22F} conducted a study on the effects of both network size and resolution in the evolution of solar metallicity massive stars. This group of studies illustrates a growing trend in the stellar evolution community to begin to quantify the uncertainties within 1D models.

Our aim in this work is to characterize the different types of CSI that may occur in massive Pop III stars and under which mixing assumptions they occur. This macrophysics impact study can serve as a guide for future targeted investigations of hydrodynamic instabilities through 3D simulations. In this way, future research may shed light on how massive stars evolve after a violent H-He interaction. For models which do not experience interactions we, follow the evolution to core collapse or onset of the pair instability. Our macrophysics impact study also reveals insights into the general conditions of H burning in Pop III stars and we can outline observable implications, for example for the C and N elemental and isotopic ratios. Section \ref{Section 2}
describes the methods used, in Sections \ref{h-burn} and
\ref{h-he} we present and discuss core and shell H burning and H-He interactions, respectively. In Section \ref{discussion} we compare our models to previous findings and discuss our results. Conclusions are given in Section \ref{conclusion}.

\section{Methods and models} \label{Section 2} 
We calculate a grid of 26 stellar models using \code{MESA} rev.\ 8118 \citep{paxton:15}. Beginning with Big Bang abundances taken from \cite{2016RvMP...88a5004C}, we explore five initial masses and mixing assumptions with the addition of one model with no convective boundary mixing, for 26 models in total. We study 15, 40, 60 ,80 and 140$\usp\Msun$ stellar models (\Tab{models}). This choice of mass range reflects our current knowledge of the Pop III initial mass function (IMF) as derived from simulations \citep{2014ApJ...792...32S,2012MNRAS.422..290S} and spans a large enough range to explore the effects of mass on both chemical evolution and H-He interactions. 

In all \code{MESA} simulations, a custom network of 151 species up to Ni was included and is shown in \Tab{net}. The JINA Reaclib database \citep{2010ApJS..189..240C} is utilized, including both strong and weak reactions for the chosen isotopes, amounting to over 1,500 reactions. We chose this network in order to realize possible proton captures from hot CNO breakout reactions as reported by \cite{keller:14} and \cite{2014ApJ...794...40T}. This network follows neutron captures no more than one isotope off the valley of stability, and only goes to the iron group therefore, if the \iprn\ occurs, we do not follow neutron capture processes realistically. Note we do not include a full Si-burning network, although this will not affect the results presented in this work.

There is a great amount of uncertainty in how much, if any, mass loss would have occurred in Pop III stars \citep{2006A&A...446.1039K}. Here we do not include mass loss, although we do allow small amounts of mass from the stellar surface to be removed if the density drops below $10^{-12}\usp\grampercc$. This is done to keep all computational cells within the given equation of state tables in this version of the \code{MESA} code. For the most part, stellar surfaces do not drop to such low densities.

For spatial resolution controls, we use \code{MESA}'s \code{mesh\_delta\_coeff}, $\delta_{mesh}$, of 0.5 during the pre-main sequence, and reduce this value to 0.2 onwards. This parameter globally refines the grid such that cell-to-cell changes in quantities such as $P$, $T$ and $\epsilon_{\mathrm{nuc}}$ are within a desired threshold, i.e., resolution is increased for regions with steep gradients in these quantities. We also apply a minimum number of total cells using \code{max\_dq}$ = 10^{-3}$, which is the maximum fraction of the total mass within each cell. This leads to an average of approximately 5,000-7,000 cells after the pre-main sequence for our models with approximately $50\--90$ cells per dex drop in H abundance at the base of the H shell. For timestepping, we similarly use a \code{var\_control\_target}, $v_c$, of $10^{-4}$ on the pre-main sequence and reduce this to $10^{-5}$ afterwards. This limits timesteps based on changes in variables such as, $\rho$, $T$ and $R$ and checks that there have not been large changes in other quantities like $L$, $X_i$ and $\epsilon_{\mathrm{nuc}}$. For changes in composition, we further limit the timesteps using the \code{dX\_nuc\_drop\_limit} controls for species with $A \leq 20$ and $X_i \geq 10^{-4}$. We choose a limiting value for \code{dX\_nuc\_drop\_limit} of  $3\ee{-3}$, which is the largest drop in mass fraction allowed due to either nuclear burning or mixing.

In H-He interactions, the convective turnover time scale and the nuclear
  time scale of H-rich material reacting with $^{12}\mathrm{C}$ from
  the He-burning shell are often similar. This regime poses a particular
  challenge to the stellar evolution code. Nuclear reactions with
  dynamically relevant energy release are operating on the same time
  scale as new fuel for these reactions is brought into the burning
  region by convection. The computationally most efficient way to
  solve the structure, mixing and nuclear network equations of stellar
  evolution is to solve each of these operators separately for each
  time step. This is a good approximation if the nuclear, mixing and
  dynamic time scales are all respectively very different. In
 H-He interactions or CSI simulations this
  operator-split approach would give inaccurate results unless the
  mixing time scale is well resolved. This is usually prohibitively
  costly and not practical. Several variations of solving two of the
  three operators jointly have been deployed. For example
  \citet{herwig:97} adopted a joint-operator solution to the nuclear
  network and diffusive mixing. While this approach leads to a
  reasonable approximation of the abundance profiles, especially the
  profile of H, it leaves out the immediate impact of the energy
  release on the structure equations.  The \code{MESA} stellar
  evolution code solves all the mixing, nucleosynthesis and structure
  equations jointly. This approach provides the most accurate 1D
  structure models for situations of nuclear energy feedback on
  convective time scales, such as in H-He interactions, as long as the fundamental
  1D assumptions of spherical and time averaging are not violated. 

As discussed in Section \ref{Section 1} for low and intermediate mass stars, it has been shown that H-He interactions can be affected by mixing assumptions. There are many different prescriptions available for a variety of extra mixing processes beyond what is included in standard MLT, and therefore, a full survey would be impractical. To limit the scope of our study we chose a sample of more commonly used prescriptions for non-rotating stars. 

We include either the Schwarzschild criterion or the Ledoux criterion for convection with some amount of additional mixing between convective layers. We employ this additional mixing in the form of convective boundary mixing (CBM) and/or semiconvection. CBM is intended to account for instabilities which occur at the boundaries of convection zones. This can include turbulent entrainment as seen in 3D hydrodynamical simulations \citep{2015ApJ...798...49W,2007ApJ...667..448M}, or internal gravity waves \citep{2003MNRAS.340..722D,2016ApJ...827...30B}.  In this work we use the exponential `overshoot' prescription given in \cite{2000A&A...360..952H} and refer to this as CBM throughout the paper. For the CBM parameter, we use either a `high' value of $f_{\mathrm{ov}}$ = 0.01 or a `low' value of  0.001. This can be compared to similar parameter studies, such as \cite{2019MNRAS.484.3921D} where $f_{\mathrm{ov}}$ ranges from 0.002 to 0.32. \cite{2020MNRAS.tmp.1745K} includes a detailed discussion of this free parameter and adopts values of $f_{\mathrm{ov}}$ = $0\--0.05$. Additionally, \cite{2019A&A...622A..50H} found that a high value of CBM ($\alpha_{\mathrm{ov}}$ = 0.05) in the form of step-overshoot was necessary to explain the extent of the main-sequence and high-end luminosity limits for red giant stars.

Semiconvection is a secular mixing process that occurs when a region within a stellar model is stable according to the Ledoux criterion and unstable according to the Schwarzschild criterion for convection. This overstable region will give rise to partial mixing as the convective velocities will be limited by the stabilizing $\mu$ gradient \citep{1958ApJ...128..348S,1970MNRAS.151...65S}. For all models using the Ledoux criterion, we include semiconvection using the prescription of \cite{1985A&A...145..179L} with an efficiency $\alpha_{\mathrm{semi}} = 0.5$. This value of $\alpha_{\mathrm{semi}}$ is within the prescribed range given in \cite{1985A&A...145..179L} and is chosen to avoid the splitting of the He burning core which occurs if the mixing speed is much lower \cite[see e.g.][]{2014ApJ...783...10S}.

 \begin{table*}
 \setlength{\tabcolsep}{3pt}
 \caption{Stellar models: Run ID, maximum central temperature during the main-sequence, maximum H shell burning temperature, main-sequence lifetime, interaction, interaction type, maximum $H-$number, and total change in mass coordinate of H-rich material. }
 \lTab{models}
 \begin{tabular}{lcccccccc}
  \hline
  Run ID$^{a}$ & $T_{\mathrm{H,core,max}}$  $(10^8\mathrm{ K})$ & $T_{\mathrm{H,shell,max}}$  $(10^8\mathrm{ K})$ & $\tau_{\mathrm{MS} \,(Myrs})$& H-He interaction & H-He interaction type$^{b}$ &$H_{\mathrm{max}}^{c}$ & $\Delta$ $\mathrm{M}/\Msun$\\                               
  \hline
  \code{15Mled} & 1.05 & 1.09 & 10.0 & $\times$ &--&   9.48\ee{-5}& 0 \\
\code{15Mledf-h} & 1.05 & 1.06 &  10.19 &$\checkmark$ & \Rshell\ &  0.05 & 0.78 \\
 \code{15Mledf-l} & 1.05 &1.03 & 10.06 &$\checkmark$ &\Rshell\ & 0.88 & 0.94 \\
   \code{15Mschf-h} & 1.05 & 1.18 & 10.22 & $\checkmark$&  \Rshell\ &  5.3\ee{-3} &0.85\\
 \code{15Mschf-l }& 1.05 &1.21 & 11.1 &$\checkmark$&  \Rshell\ & 0.11 & 0.91  \\
  
  \code{40Mled} & 1.27 & 1.58 & 4.46 &$\checkmark$& \Cshell\  & 0.22 & 1.28 \\
\code{40Mledf-h} & 1.27 & 1.06 & 4.53 & $\checkmark$& \Rcore\  & 8.59\ee{-5} & 2.48 \\
 \code{40Mledf-l} & 1.27 &1.29 & 4.50  &$\checkmark$& \Rshell\  & 1.97\ee{-6} & 0.26\\
   \code{40Mschf-h} & 1.27 & 1.23 & 4.63 & $\checkmark$ & \Rshell\ &  0.35 & 2.80\\
  \code{40Mschf-l} & 1.27 & -- & 5.32 &$\checkmark$& \Rcore\ &  0.62  & 2.24\\

   \code{60Mled} & 1.33& 1.89  & 3.52 &$\checkmark$& \Cshell\ &  0.28 & 0.40\\
 \code{60Mledf-h} & 1.33 & 1.44 & 3.56 & $\checkmark$&  \Rshell\ & 0.74  & 1.77\\
 \code{60Mledf-l} & 1.33 &1.28 & 3.53 & $\times$ &--  &  5.9\ee{-4} & 0 \\
  \code{60Mschf-h} & 1.33 &-- & 3.69 & $\checkmark$& \Ccore\ &1.3\ee{-3} & 23.24$^{d}$ \\
 \code{60Mschf-l} & 1.33 & --& 3.65 &$\checkmark$& \Ccore\ & 1.3\ee{-3}& 22.91$^{d}$ \\
  
   \code{80Mled} & 1.36 & 2.34 & 3.07 &$\times$& -- & 4.05\ee{-4}& 0\\
\code{80Mledf-h} & 1.36 & 2.23 & 3.10 & $\checkmark$ &  \Rshell\ & 0.74 & 3.12\\
 \code{80Mledf-l} & 1.36 & 1.28 & 3.07 &$\checkmark$ &  \Rshell\ & 1.82\ee{-6}  & 0.39 \\
  \code{80Mschf-h} & 1.36 &-- & 3.37 & $\checkmark$& \Ccore\  & 0.07 & 35.6 \\
 \code{80Mschf-l} & 1.36 & 1.44 & 3.46 &$\checkmark$& \Rshell\ &  2.42e-05 &  0.16  \\
  
  \code{140MNoMix} & 1.42 &-- & 2.71 &$\checkmark$& \Ccore\ & 0.01 & 59\\
  \code{140Mled} & 1.42 &  2.18  & 2.48 &$\times$& -- &  7.27\ee{-4} & 0 \\
 \code{140Mledf-h} & 1.42& -- & 2.50 &$\checkmark$&  \Rcore\ & 0.32& 6.05 \\
  \code{140Mledf-l }& 1.42 & 1.36 & 2.49 &$\checkmark$&  \Rshell\ &  1.64\ee{-7} &1.45\\
   \code{140Mschf-h }& 1.42 &--& 2.76 &$\checkmark$& \Ccore\ & 0.01 & 73.61$^{d}$\\
  \code{140Mschf-l }& 1.42 & --& 2.74&$\checkmark$& \Ccore\ & 3.05\ee{-5} & 71.21\\
  \hline
 \multicolumn{8}{l}{$^a$ Run IDs represent the mass, criterion for convection, and boundary mixing parameters used, e.g. \code{15Mledf-h} is a $15\Msun$ model with the }\\
\multicolumn{8}{l}{Ledoux criterion for convective stability with semiconvection and our larger value of CBM. See section \ref{Section 2} for details on the models.}\\
\multicolumn{8}{l}{$^b$ Naming used refers to either a convective (C) or radiative (R) He-rich shell or core region that the convective H shell is interacting with. }\\
  \multicolumn{8}{l}{$^c$ $H$ as defined in \Eq{1}. }\\
   \multicolumn{8}{l}{$^d$ See Section \ref{C core He} for a description of \code{60Msch-} and \code{140Mschf-h} interactions.}  \\
 \end{tabular}
\end{table*}
 
 \begin{table}
 \centering
 \caption{Nucleosynthesis network used in \code{MESA} simulations. }
 \lTab{net}
 \begin{tabular}{lcccc}
  \hline
  Element & Z  & Element & Z  \\                             
  \hline
  H & 1-2 &  P & 27-32  \\   
  He & 3-4  &  S & 29-37  \\   
  Li & 6-7 &  Cl & 31-40 \\
  Be & 7-9  & Ar & 33-42\\
  B & 10 &  K & 36-43   \\
  C & 11-13  &  Ca & 37-48 \\
  N & 12-15  & Sc & 40-45 \\
  O & 14-18  &  Ti & 41-53  \\
  F & 16-19  &  V & 47-52  \\ 
  Ne & 18-22  &  Cr & 49-54  \\      
  Na & 20-23  &   Mn & 53-55  \\   
  Mg & 21-26 &   Fe & 54-58  \\    
  Al & 23-27 &   Ni & 56  \\     
  Si & 25-31  &   &   \\      
 \end{tabular}
\end{table}
\section{Hydrogen burning} \label{h-burn}
\subsection{Core H burning} \label{corehburn}

Given their initial primordial composition, massive Pop III stars begin H-burning via p-p chain reactions. These reactions are energetically insufficient to maintain complete equilibrium at temperatures over $\approx 1.5 \ee{7} \unitstyle{K}$, where CNO is the dominant energy source above these temperatures in higher metallicity stars. The star contracts until temperatures are high enough ($\sim$10$^8\usp\K$) to initiate 3$\alpha$ reactions, thus producing  a $^{12}\mathrm{C}$ mass fraction of  $\sim 10^{-10}$---sufficient to activate the CNO cycle \citep{1971Ap&SS..14..399E}. The time it takes for CNO to activate is inversely proportional to the initial mass of the star, with less massive stars having more of the main-sequence (MS) lifetime supported by p-p reactions and more massive stars generating the C necessary to supply CNO much earlier. This can be seen in \Fig{schftcrhoc} where red dots indicate the time at which CNO overcomes p-p chains as the dominant source of nuclear energy generation in the core. This has very important consequences for the structure of the star. It is overall more compact and the temperature for H-burning becomes closer to that of He-burning. This in turn leads to a smaller temperature difference between H- and He-burning layers than would be seen at higher metallicities. In \Fig{40M_sch_f_kip} around $5.3\usp\unitstyle{Myr}$ this manifests itself as a smooth transition from H- to He-core burning.

\begin{figure}
	\includegraphics[width=\columnwidth]{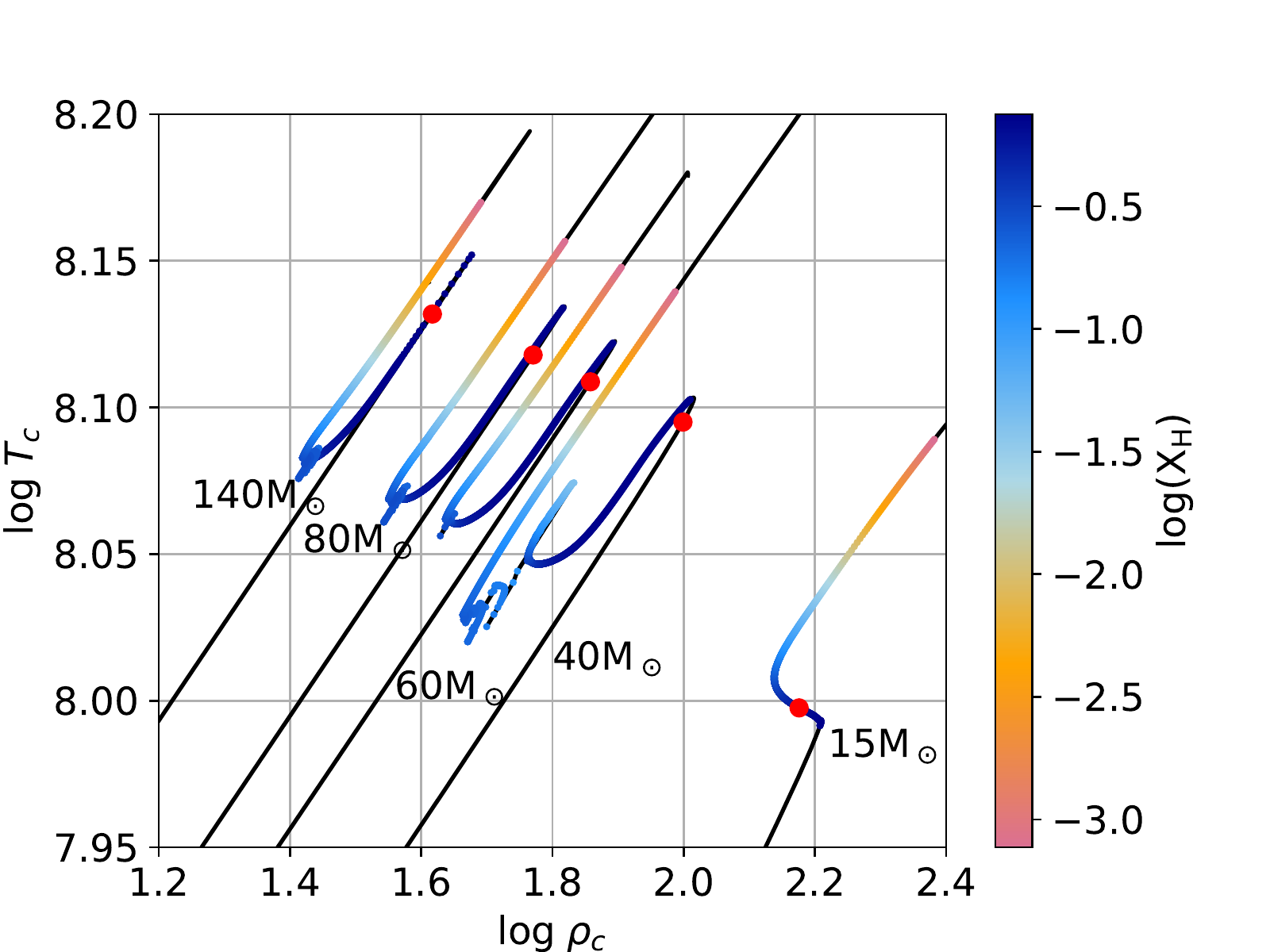}
    \caption{$\rho_\mathrm{c}- T_\mathrm{c}$ diagram for \code{schf-h} models using the Schwarzschild criterion for convection and our higher adopted value of CBM, where  mixing of material into the H core occurs during the MS phase in models $\geq 40\usp\Msun$ (see  \ref{corehburn}). Colours indicate the logarithm of the mass fraction of H remaining in the centre of the star. Red dots show the point in evolution at which CNO reactions overtake p-p chains as the dominant energy source in the centre.}
   \lFig{schftcrhoc}   
\end{figure}

\begin{figure}
	\includegraphics[width=\columnwidth]{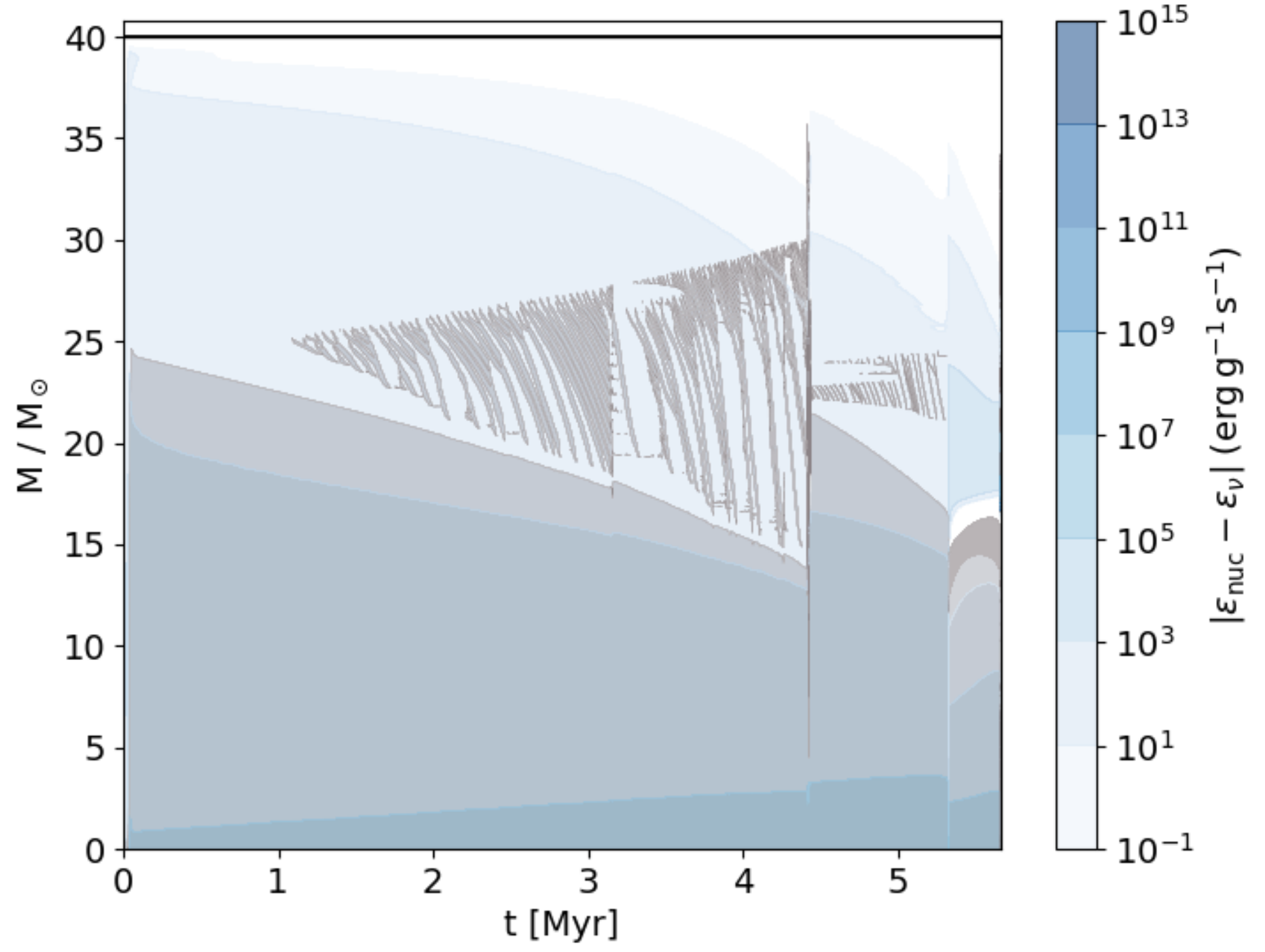}
    \caption{Kippenhahn diagram of the MS and beginning of He-burning in the \code{40Mschf-h} model with the Schwarzschild criterion for convection and $f_{\mathrm{ov}}$ = 0.01. Gray areas indicate convection, blue contours show specific energy generation. An MS mixing event can be seen at $ t \approx 4.4 \usp\unitstyle{Myr}$. }
   \lFig{40M_sch_f_kip}
\end{figure}

In all of our simulations there are small convection zones above the H-burning core. These small convective `fingers' present early in the MS phase---as soon as the H core begins receding---and descend downward during the MS (e.g. \Fig{40M_sch_f_kip} above the convetive H core). These convective regions above the H-burning core are the result of composition changes induced by the receding core which ultimately produces a staircaising effect on the composition profiles \citep[see, e.g.][]{1958ApJ...128..348S,1985A&A...145..179L, 2000ApJ...528..368H}. There is a small amount of p-p chain burning activity in these regions. 

In simulations with masses $\geq 40 \usp\Msun$, these convective fingers penetrate into the receding H core, injecting fresh hydrogen fuel from above. They are found in \Fig{40M_sch_f_kip}, above the H core, and in all other Kippenhahn diagrams presented in this work.
The most extreme examples found in our simulations are those with greater CBM and without the regulating effects of semiconvection---the \code{-schf-h} cases. This can be seen in \Fig{schftcrhoc} for all masses $\geq 40 \usp\Msun$ near the minimum  $\rho_\mathrm{c} \-- T_\mathrm{c}$ value on the MS. In the \code{40Msch-f} model, within the span of $\approx30,000\usp\yr$ the mass fraction of H increases from 0.05 to 0.17 and the convective core expands both in mass and radial coordinates. 30,000$\usp\yr$ later, more mixing occurs and $X_{\mathrm{H}}$ increases to 0.25, further extending the MS lifetime, which can be seen at $t \approx 4.3\usp \unitstyle{Myr}$ in \Fig{40M_sch_f_kip}. This also causes a sudden increase in the luminosity. Similar behaviour was described by \cite{2016ApJS..227...22F, 2018ApJ...860...93S, 2020MNRAS.tmp.1745K}. \cite{2016ApJS..227...22F} found this in a solar metallicity model with an initial mass of 30$\usp\Msun$ using \code{MESA}. They found that this event appeared when using a high mass resolution but not their adopted low-resolution. Their adopted high and low mass resolutions for this portion of the study are $\Delta M = 0.01$ and $0.02\usp\Msun$ per cell, respectively, and also include models with a maximum mass resolution of 0.005$\usp\Msun$ per cell. For comparison, our mean mass resolution on the MS of the \code{40Msch-f} model is 0.008 $\usp\Msun$ per cell.  \cite{2020MNRAS.tmp.1745K} found the presence of convective fingers decreased in frequency with $f_{\mathrm{ov}} \gtrsim 0.01$ and did not occur in stars of $15\usp\Msun$ at or above approximately this value. In our models, we find convective fingers in all of our simulations although core penetration does not happen in our $15\usp\Msun$ models. The difference in outcomes may relate to the differences in opacity and/or temperature gradients in Pop III vs. solar metallicity stars and/or our smaller range of CBM values.  

While CBM itself will supply fresh H and increase the overall MS lifetime, the drastic H-mixing events in the models presented here can lead to MS lifteimes up to $20\%$ longer relative to simulations of the same mass without them (see \Tab{models}). This is greater than the $\approx 5\%$ increase reported by \cite{2016ApJS..227...22F} which may be due to the difference in our maximum CBM parameter of 0.01 and their value of 0.001. Moreover, in our simulations the convective fingers above the H core are present in most models using CBM but is less extreme when semiconvection and the Ledoux criterion are used. These relevant physics choices are used by \cite{2016ApJS..227...22F} in all simulations. While including Ledoux and semiconvection at this evolutionary stage appears more physical, we do not find there to be any impact from these dramatic MS mixing events on the behaviour of later H-He interactions and regard them as numerical artefacts. 

A comparison of the \code{15Mschf-h} and \code{40Mschf-h} models indicates that this mixing on the MS may be related to the overall compactness, temperature gradients above the H cores and the associated nuclear feedback. More specifically, a $15\usp\Msun$ Pop III star has lower central temperatures than its more massive counterparts, and thus, more time on the MS spent in p-p chain burning ($17\%$ vs $0\%$) and lower energy generation from CNO reactions. Simultaneously, the $15\usp\Msun$ simulation has nearly 2 times higher core and overall densities during the MS. Overall, the \code{15Mschf-h} model is a more compact star with steeper entropy gradients which may act to inhibit the mixing on the MS seen in higher mass models.

\subsection{Shell H-burning} \label{shellhburn}
H-burning products that will carry over to subsequent stellar generations are produced primarily in H-burning shells. In Pop III stars, H-burning shells tend to be hotter than in their higher metallicity counterparts for reasons presented in Section \ref{corehburn}. All simulations presented in this work have convective H-burning shells for at least a portion of the post He-core burning phase. 

\Tab{models} lists the maximum H-shell burning temperatures found in our grid of models. This is the maximum temperature at the lowest mass coordinate where the H mass fraction is at least $10^{-4}$. If there is a H-He interaction, the maximum H-burning temperature reflects how far into the He shell the H-rich material descends by the end of the simulation. This distance the H-burning shell has advanced into the He-rich region, $\Delta M / \Msun$, is also reported in the final column of \Tab{models}.  

In simulations over $15\usp\Msun$, the H-shell temperature increases over time. The \code{15Mled} and \code{15Mledf-l} runs both have the maximum temperature at the beginning of H-burning and decrease by about $10^7\usp \unitstyle{K}$ by the end of the simulation. Otherwise, the maximum temperatures are found either at the end of the simulation in those that do not have an H-He interaction, or during a H-He interaction. For the \code{140Mled} run, the maximum temperature of the H shell is measured prior to encountering the pair instability, as is described below in Section \ref{None}.

H-He interactions can have significant effects on the physical conditions and nucleosynthesis within H-burning shells. The details of the nucleosynthesis are described in Section \ref{h-he} and Section \ref{CN}.

\subsection{Hot CNO}\label{hotcno}
The activity of hot CNO cycling, and possible breakout therefrom, may have important implications for the abundances found in the second generation of stars. Core- and shell-H burning conditions can be investigated relating to CNO breakout in the four models that do not experience a H-He interactions. 

The first hot CNO cycle occurs when protons preferentially capture on the $^{13}\mathrm{N}$ nucleus before it can decay into $^{13}\mathrm{C}$, which begins at temperatures of $\sim 1\ee{8}\usp\unitstyle{K}$ \citep{2010ARNPS..60..381W} for sufficiently high densities. This leads to the bypass of the $^{13}\mathrm{N}\ \beta^{+}$ decay, and energy generation is controlled by the $ \beta^{+}$ decays of $^{14}\mathrm{O}$ and $^{15}\mathrm{O}$, the slowest reactions in the cycle. During core contraction on the MS in each of our four simulations without a H-He interaction, the conditions become hot and dense enough such that hot CNO conditions are temporarily met. This is indicated by the production of $^{14}\mathrm{O}$ from the $^{13}\mathrm{N}(p,\gamma)$ reaction, with a lifetime, $\tau_{p}$, that becomes shorter than the beta decay lifetime. Hot CNO conditions last for $\sim 1\%$ of the total MS lifetime and $< 1\usp\unitstyle{yr}$ at most during H-shell burning in our simulations. Therefore, cold CNO dominates over hot CNO for both core and shell H burning in the \code{15Mled}, \code{80Mled},  \code{140Mled}, and  \code{60Mled-l} simulations.  

While hot CNO cycles are effectively closed at these temperatures \citep{2007nps..book.....I}, all our \code{MESA} simulations show small amounts of elements with charge $Z\geq 10$, being produced during H burning. The greatest of these is Ca, owing to its doubly magic nature, with an average mass fraction of no greater than about $10^{-12}$ in the H-rich envelope. 98\% of the Ca in the H-burning shell in the four simulations we examine is produced during the MS. \Tab{models} shows our highest H-burning temperature is found in the \code{80Mled} model where the peak temperature is reached during the last day of simulation time. \Fig{80Mabu} shows the most abundant elements and $\mathrm{Ca}$ for the final time step. All of the simulations presented in this work have convective H-burning shells. Post core-He burning in the \code{80Mled} model, the H-rich convection zone extends from $\approx 35\usp\Msun $ to $\approx70\usp\Msun$. After core-O burning, this convection extends nearly to the surface of the model, bringing with it any H-burning material present.

\begin{figure}
	\includegraphics[width=\columnwidth]{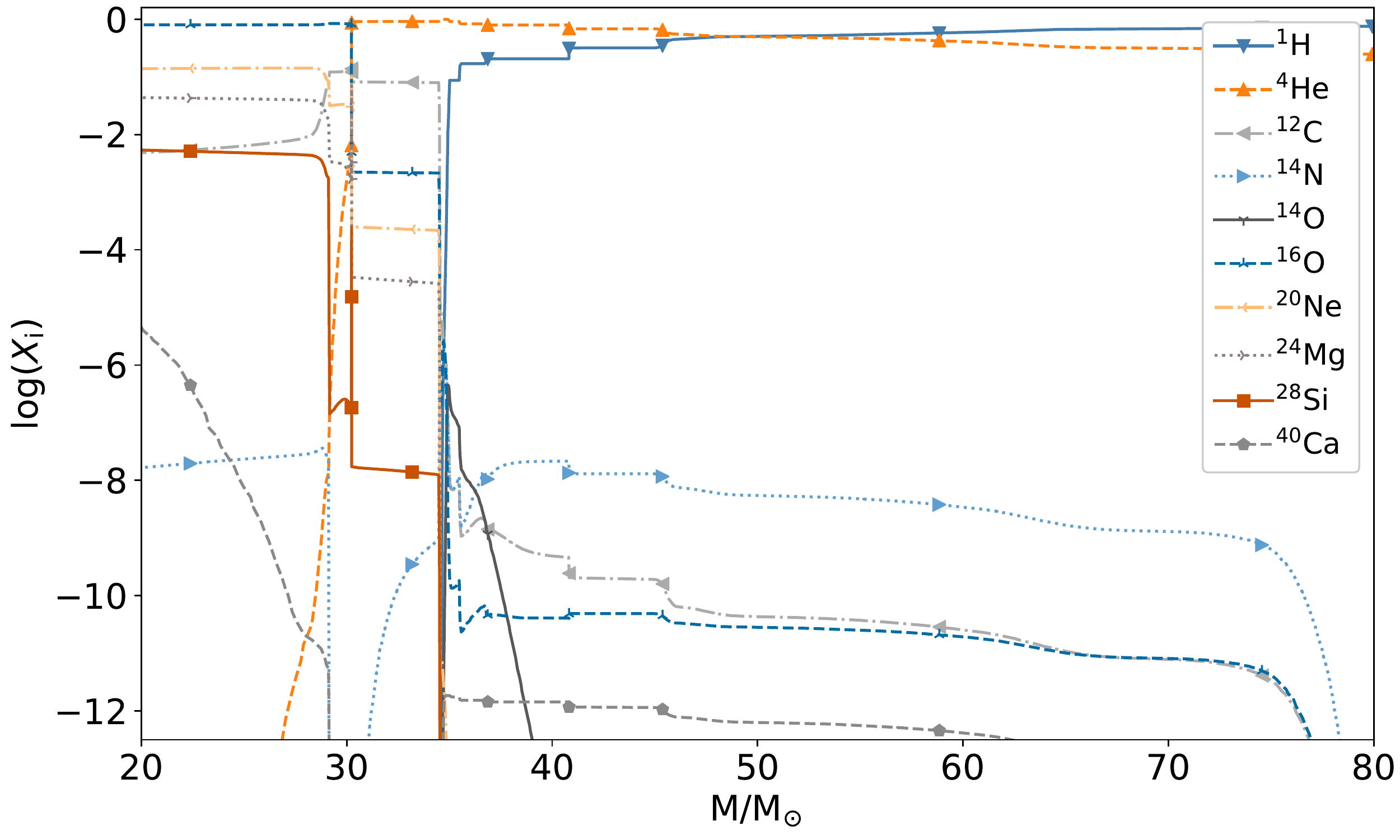}
    \caption{Mass fractions of the most abundant species and Ca for the final model of the \code{80Mled} model with the Ledoux criterion and semiconvection. Plot extends from the middle of the C-shell to the surface.}
   \lFig{80Mabu}
\end{figure}

Hot CNO breakout reactions are reported by \cite{keller:14, 2017A&A...597A...6N} and \cite{2014ApJ...794...40T} as the production site for the Ca in Pop III stellar evolution models. Those yields have been used to match the abundances observed in Ultra Metal-Poor Cemp-no stars. \cite{2018MNRAS.474L..37C} found that the \iprn\ was the primary source for Ca in their simulations.

 \begin{table}
 \centering
 \caption{Single-zone calculation parameters.}
 \lTab{ppn}
 \begin{tabular}{lcc}
  \hline
  Description & $T \usp (10^8\mathrm{ K})$  & $\rho \usp (\grampercc)$\\                             
  \hline
  \code{80Mled} H-core average & 1.19 & 39.8  \\
  \code{80Mled} H-core maximum & 1.36 & 65  \\
  \code{80Mled} H-shell maximum & 2.34 & 1.77  \\
 \end{tabular}
\end{table}

To determine the sequence of breakout reactions and reactions leading to the production of Ca, we have run three constant temperature and density single-zone calculations using the \code{PPN} code \citep{2016ApJS..225...24P}. These calculations use the same network as described in \cite{2018MNRAS.474L..37C} but are run with the time-weighted average temperature for core H burning, maximum temperature for core H burning, maximum temperature for H-shell burning, and corresponding densities. The temperature and densities used are given in \Tab{ppn}. 

For both of the core H burning \code{PPN} runs, we find that the small amount of breakout that occurs, flows primarily through the $^{16}\mathrm{O}(p,\gamma)^{17}\mathrm{F}(\beta^+ \nu)^{17}\mathrm{O}(p,\gamma)^{18}\mathrm{F}(\beta^+ \nu)^{18}\mathrm{O}(p,\gamma)^{19}\mathrm{F}(p,\gamma)^{20}\mathrm{Ne}$ with a small contribution from proton captures and subsequent decays on $^{17}\mathrm{F}$ and $^{18}\mathrm{F}$. Here it should be noted that the flux, $\mathrm{d}Y_i/\mathrm{d}t$, of the $^{18}\mathrm{F}(p,\alpha)^{15}\mathrm{O}$, $^{18}\mathrm{O}(p,\alpha)^{15}\mathrm{N}$ and $^{17}\mathrm{O}(p,\alpha)^{14}\mathrm{N}$ reactions are $\approx  10 \--4,000$ times greater than competing breakout reactions and decays under these conditions. More importantly, the flux through the $^{19}\mathrm{F}(p,\alpha)^{16}\mathrm{O}$ is $\approx 5,500$ times greater than through the competing $^{19}\mathrm{F}(p,\gamma)$ channel, keeping most material within the CNO.  For the small amount of material that can make it past F, proton captures and decays along the valley of stability lead to a small production of Ca with mass fractions no greater than $10^{-12}$.

Investigating the maximum H-shell temperatures from our \code{80Mled} simulation, we find a somewhat different story. In addition to the aforementioned reactions, $^{15}\mathrm{O}(\alpha,\gamma)^{19}\mathrm{Ne}$ occurs at these temperatures, although the competing flux through the decay of $^{15}\mathrm{O}$ is $\sim 10^{7}$ times greater, making its contribution completely negligible. As in the previous two simulations described, material passes through $^{17}\mathrm{F}(\beta^+ \nu)$ and $^{18}\mathrm{F}(p,\gamma)^{19}\mathrm{Ne}$, but at these higher temperatures, some is able to bypass $^{19}\mathrm{F}$, when decays and proton captures lead this flow to $^{21}\mathrm{Ne}$ and $^{23}\mathrm{Na}$. The fluxes through breakout reactions we have discussed are in the range $\sim \ee{-23} \-- \ee{-18}\unitstyle{s^{-1}}$. Any occurring hot CNO breakout during shell H-burning has little impact on the Ca abundance due to the limited time prior to collapse. \cite{2014ApJ...794...40T} reported temperatures over $4.5\ee{8}\usp\unitstyle{K}$ leading to hot CNO breakout in a $140\usp\Msun$ simulation during shell H burning. We cannot confirm this result as the single $140\usp\Msun$ simulation we have without a H-He interaction, the \code{140Mled} case, becomes pair-unstable. The maximum temperature in the H shell before the instability occurs is $2.18 \ee{8}\usp\unitstyle{K}$. 

To determine wether this Ca production can explain the Ca abundances in the most Fe-poor stars, we compare the [Ca/H] value of -6.94 given in \cite{2017A&A...597A...6N} for the most Ca-poor star currently known, SMSS 0313-6708 \citep{keller:14} with calculated values from \code{MESA} simulations presented here. Performing a simple calculation, we use solar abundances from \cite{2009ARA&A..47..481A} and assume a faint supernova. We calculate abundances using:

  \begin{equation} 
  \label{eq2}
X_{\mathrm{total}} = \frac{\int_{M_{\mathrm{B}}}^{M_{\mathrm{tot}}} {X} dm}{\int_{M_{\mathrm{B}}}^{M_{\mathrm{tot}}}{dm}} 
 \end{equation}

Where $X$ is the mass fraction of a given isotope, $M_{\mathrm{tot}}$ is the mass coordinate corresponding to the surface of the star, and $M_{\mathrm{B}}$ is the mass coordinate corresponding to the base of the H envelope. We do not include nucleosynthetic contribution from the explosion or any subsequent mixing. We also do not include any dilution from the ISM and therefore, our estimates represent the maximum [Ca/H] value that can be obtained by these simulations given the aforementioned assumptions. We find [Ca/H] values of -7.92, -7.72 and -8.91 for the final models of our \code{80Mled}, \code{60Mledf-l} and \code{15Mled} simulations, respectively. Therefore, the observed Ca abundance is between $\sim 0.8$ and nearly  $2\usp\unitstyle{dex}$ larger than the largest amount of Ca made by hydrostatic H burning in these models. 

If the branching ratio for $^{19}\mathrm{F}(p,\gamma)/ ^{19}\mathrm{F}(p,\alpha)$ were a factor of $\approx 10$ higher, this would help relieve the tension. In our \code{PPN} simulation using the maximum shell temperature from the \code{80Mled} model, increasing the $^{19}F(p,\alpha)$ rate by a factor of 10 results in an increase in the Ca abundance of $1\usp\unitstyle{dex}$. Though recent measurements of the $^{19}F(p,\alpha)$ rate \citep{2019PhRvC.100d4307L} and investigations into $^{19}F(p,\gamma)$ \citep{2008PhRvC..77a5802C}, which have not been adopted in this work, both suggest a lower branching ratio than used in this work by at least a factor of 10, which we expect to lower the abundance of Ca created in simulations.

\section{H-He interactions} \label{h-he}

We investigate the occurrence and behaviour of convective H-He interactions in our stellar evolution models. \Tab{models} lists which type of interaction occurs in each simulation. 22 out of 26 simulations have H-He interactions. In all cases the H-burning layers are convective. The H-burning convective shell can interact either with core or shell He burning. The He-burning core or shell can be either radiative (\texttt{Rad}) or convective (\texttt{Conv}). In our simulations we find examples of all of these four combinations, which we label \Rshell\ for H-shell material mixing with a radiative He-shell, \Cshell\ for a convective He-shell, and likewise \Rcore\ and \Ccore\ for H-rich material interactions with the radiative region above the He-burning core or with the convective core itself, respectively.

We define H-He interactions as a region where $X_{\mathrm{H}}$ increases rapidly (over several time-steps) to $> 10^{-4}$ due to mixing over any radial portion of a He-rich layer. This choice is made to properly recover H-He interaction properties. When protons are being mixed into convective He layers they are transported downward and able to react with the ashes of the He-rich fluid. For example, in the \code{80Mscf-h} case, described in Section \ref{C core He}, where protons are mixed into a convective He core, energy generation from CNO reactions dominates over that from triple-$\alpha$ reactions by several orders of magnitude if the H mass fraction is at least $> 10^{-4}$. Therefore, we use this value for consistency in all calculations and measurements requiring the location of the `base' of the H-rich material.

We describe the evolution of four models in detail, one for each mode listed in \Tab{models}, and briefly describe other simulations of note. In order to gain insight into the energetic feedback from nuclear reactions during a H-He interaction, we adopt the $H$ number as given in \cite{2016MNRAS.455.3848J} defined as:
\begin{equation}
H = \epsilon_{\mathrm{nuc}} \tau_{\mathrm{conv}} / E_{\mathrm{int}}
 \lEq{1}
\end{equation}

This number is an estimate of the energy being deposited into the flow from nuclear reactions relative to the specific internal energy. The convective advection timescale, $\tau_{\mathrm{conv}}$, is the shortest timescale over which energy can be transported by convection. This is calculated as the pressure scale height at the base of the convective H-rich layer divided by the maximum local convective velocity. This is then compared to the local internal energy, $E_\mathrm{int}$. For a general equation of state, $ \phi E_\mathrm{int} =-E_\mathrm{g}$. For the case of an ideal gas $ \phi =2$, and for a photon gas $ \phi =1$. As with most massive stars, massive Pop III stars have a non-negligible radiation contribution to their equation of state. In the absence of other considerations, this shows that it becomes increasingly easy to unbind material with an increasing contribution of radiation to the total pressure. Therefore, the $H$ number can be interpreted as an indication of how far from hydrostatic equilibrium the simulation is. It should be noted that when these numbers are calculated, many models have already violated the assumptions of MLT, and $H$ numbers are to be taken as instructive. In general, convective shell interactions with larger $H$ numbers are in principle more likely to lead to ejections compared to cases with lower $H$ number. However, the 3D fluid-dynamics of the event would be an additional determining factor. For example, dynamic instabilities such as the global oscillation of shell H-ingestion (GOSH) reported by \cite{2014ApJ...792L...3H} could lead to focused non-spherical ejections into a confined solid angle. This would require only a fraction of the binding energy of the complete $4\pi$ shell of mass $\Delta M$ for successful partial ejection. Given these uncertainties, we refrain from making specific predictions regarding which $H$ numbers will lead to mass ejection.

When H-He interactions begin, time steps rapidly decrease and the simulations fail to converge. Up to this point, the models describe the sequence of events leading up to these interactions. We do not include any `stellar engineering' parameter modifications to continue the evolution through the violent phases, as in most cases the assumptions of 1D stellar evolution are no longer appropriate, for example in the case of high $H$ number.  

To determine whether convective-shell interactions were caused, as opposed to influenced, by any type of boundary mixing we ran one simulation, \code{140MNoMix}, that uses the strict Schwarzchild criterion. Also in this case a \Ccore\ interaction is observed. 

Some commonalities can be found in these simulations not specific to any of the cases described below. For example, H-He interactions are initiated during core contraction phases from the beginning of core He burning to the onset of core O-burning. We also find that H always moves downward first, but in many simulations C and other He-burning products move upward into the H shell as well. This can boost mass fractions of CNO elements, having implications for the distribution of observed  $^{12}\mathrm{C}/^{13}\mathrm{C}$ ratios in second generation stars \citep{2017A&A...605A..63C}. Energy generation and the $H$ number tend to depend on the amount of He burning that has occurred in the He-rich layer, as this dictates the $^{12}\mathrm{C}$ abundance which interacts with the protons from the H-shell. 

\subsection{Convective H shell and radiative He shell}\label{R shell He} 
\Rshell\ interactions take place in nearly half of the models presented in this work. This interaction type takes place sometime after core He-burning and in it, a convective H-burning shell descends downward into a previously H-free radiative He burning shell. Notably, this interaction type is only seen in models with CBM included.

\begin{figure}
	\includegraphics[width=\columnwidth]{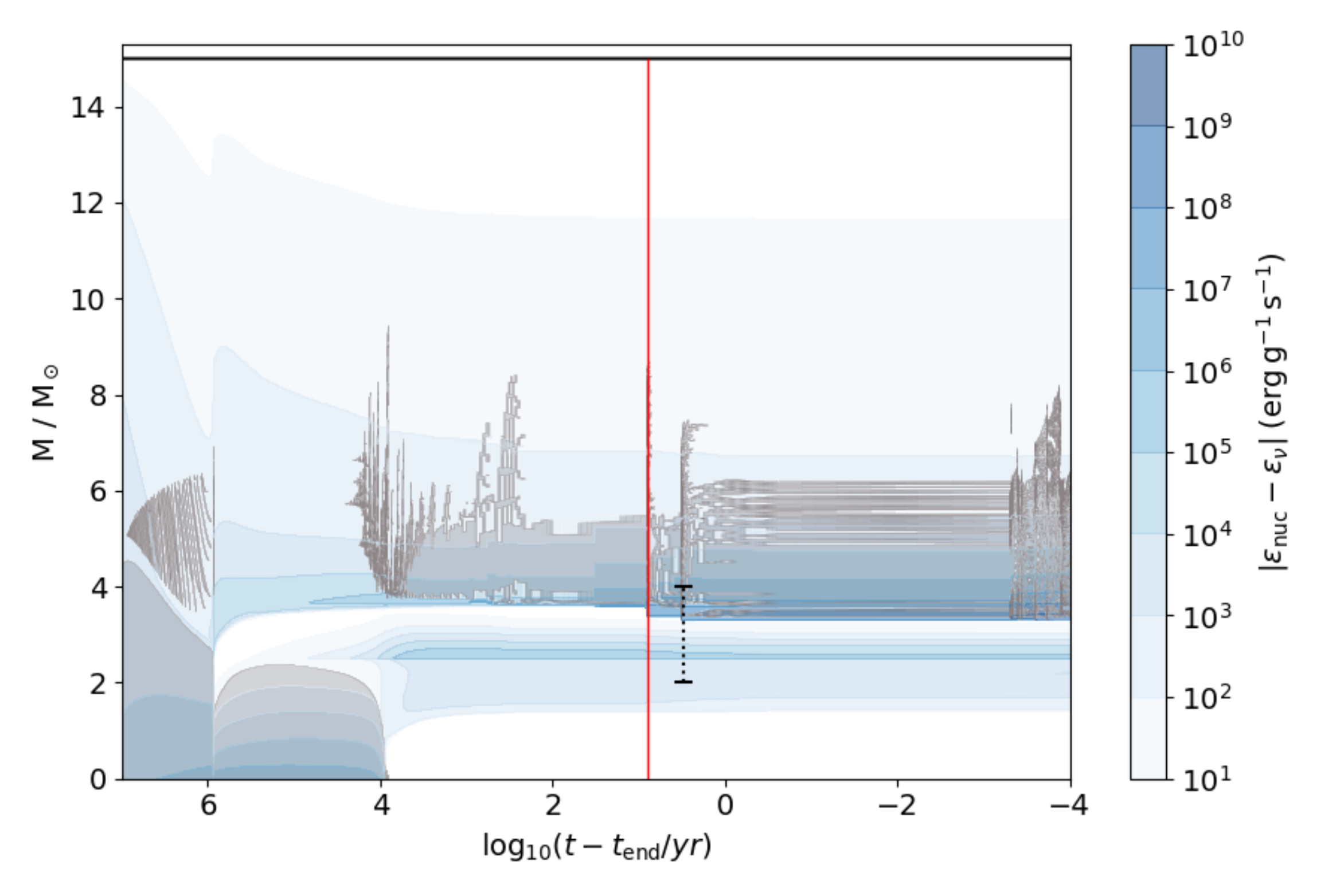}
    \caption{Kippenhahn diagram of the total calculated evolution of the \code{15Mschf-h} \Rshell\ case, using the Schwarzschild criterion for convection and $f_{\mathrm{ov}}$ = 0.01. Colours are the same as in \Fig{40M_sch_f_kip}. The red line indicates the beginning of the H-He interaction and the black dotted line indicates where the profiles in \Fig{15Mother} are taken.}
    \lFig{15Mschf}
\end{figure}

\begin{figure*}
	\includegraphics[width=\textwidth]{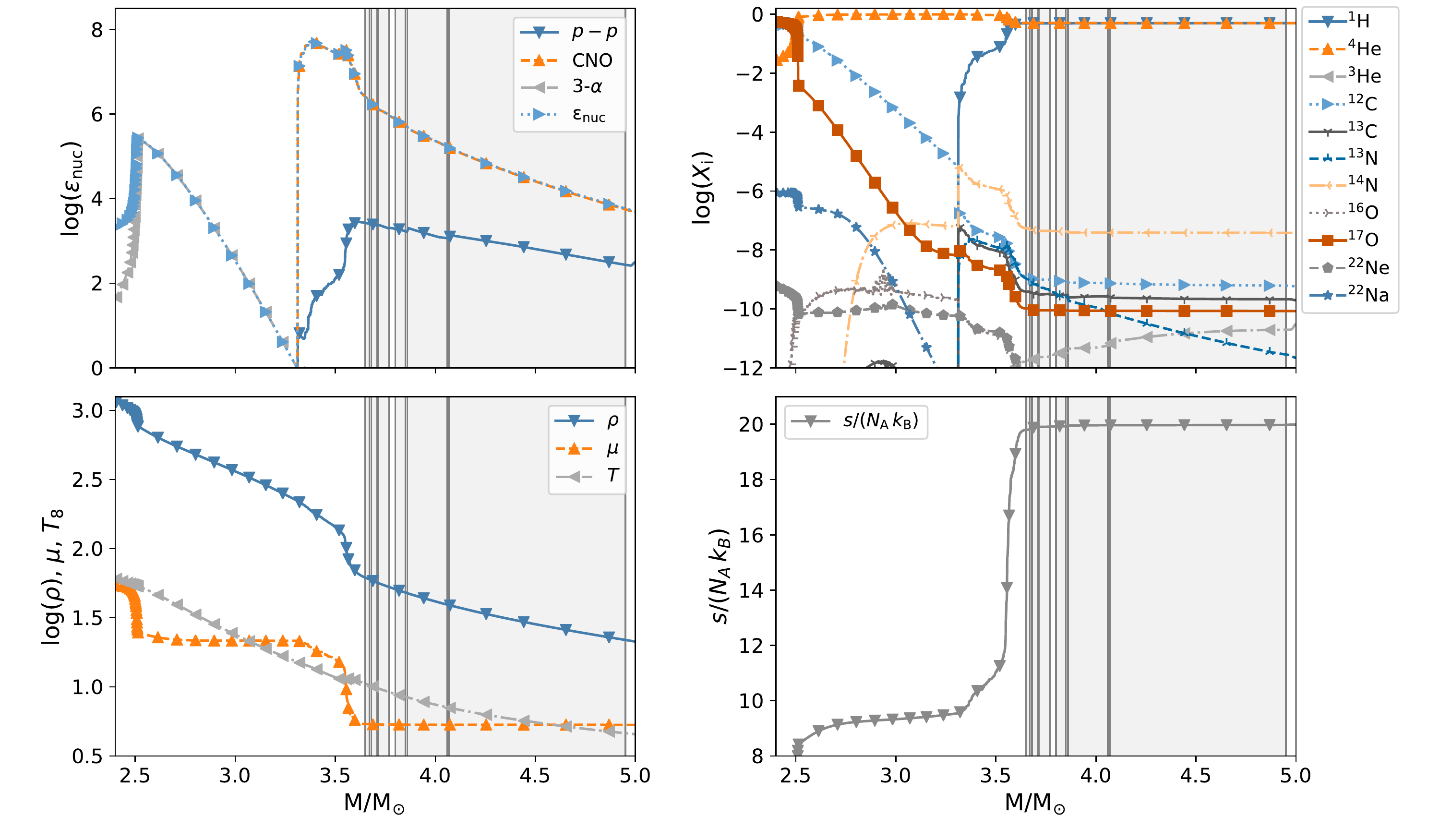}
    \caption{\code{15Mschf-h} as an example of an \Rshell\ He interaction. \textbf{Top left}: Total specific energy generation and that from CNO, p-p and tri-$\alpha$ reaction groups. \textbf{Top right}: Mass fractions of several abundant species. \textbf{Bottom left}: Temperature ($T_8 \equiv T$/$10^{8}\usp\K$) , mean molecular weight and density profiles. \textbf{Bottom right}: Specific entropy. Profiles taken $\approx 5 \usp\unitstyle{yr}$ after H-He interaction begins, times for which are indicated in \Fig{15Mschf} as black dashed and solid red lines, respectively. Grey areas show regions unstable to convection. See Appendix \ref{appendix} for profiles of diffusion coefficents}.
    \lFig{15Mother}
\end{figure*}

We use the \code{15Mschf-h} simulation as a representative case for this type of interaction. \Fig{15Mschf} shows the total evolution of the simulation until $\sim 5\usp \unitstyle{min}$ before the simulation ends. As with all the $15\usp\Msun$ simulations, convective mixing into the H core, as described in Section \ref{h-burn}, does not occur on the MS. At the end of core He burning the inner regions of the star contract to initiate core C burning, creating a convective H shell atop a radiative He shell. As the H shell begins developing, several small, convectively unstable regions merge, spanning over $1\usp\Msun$ in Lagrangian space. As this happens, the He shell temperature increases and energy generation from triple-$\alpha$ reactions causes the region to expand and push toward the H-burning convection zone. Simultaneously, over the course of $\approx 2\ee{4}\usp\yr$ the H shell descends inward until it reaches the top of the previously H-free core (at $\mathrm{log(t-t_{end})} \approx 3.5$ in \Fig{15Mschf}). The effective CBM region at the base is the H shell is during this time is $\approx 0.07\usp\Msun$. Small amounts of protons begin to mix at the top of the expanding He shell, and C and O are mixed upwards into the H-shell. Abundances of $^{12,13}$C, $^{13,14}$N and$^{16}$O increase in mass fraction by about $1\usp\unitstyle{dex}$ within $1000\usp\unitstyle{yrs}$. The addition of catalysts to the H-rich material creates non-equilibrium CNO abundances which are nearly restored within 6 months. This upset and partial restoration to equilibrium continues for the $8\usp\yr$ duration of the simulation as new $^{12}$C is introduced to the H-rich material. \Fig{15Mother}\footnote{Note that grey regions in all panel plots of this type throughout this paper show the approximate location of convection zones. Once a H-He interaction has begun, convective regions as reported by \code{MESA} often break up into many small, frayed zones. Therefore, we plot the innermost and outermost points of the largest zones. For profile of diffusion coefficients See Appendix \ref{appendix}} also shows that within about $5 \usp\yr$, much of the  $^{12}$C which was mixed into the H shell has been converted to $^{14}$N, and this  $^{14}$N is also mixing down into the He shell. Elements heavier than O are not produced in the H-burning layer at mass fractions $ > 10^{-12}$. With both values of CBM this interaction type happens in 15\Msun\ simulations using both convection criteria. H-He interaction does not occur for this mass in models without any CBM. For a timespan of 600\usp\yr, ingression of the proton rich material into the previously H-depleted radiative layer reaches down $0.06\usp\Msun$. This creates a small shelf in the entropy profile likely due to the change in mean molecular weight and CNO energy generation, as can be seen in the early stages at a $M/\Msun \approx 3.5$ in \Fig{15Mother}. The convective H-rich material then penetrates downward in mass coordinate to $3.34\usp\Msun$ and then $3.3\usp\Msun$ within $3.2\usp\yr$ before the simulation stops. In this time, the specific energy generation from CNO in the region has increased from $\sim 10^6\usp \ergspergs$ prior to the event to $10^{8.6}\usp \ergspergs$ at the final model.

\begin{figure}
	\includegraphics[width=\columnwidth]{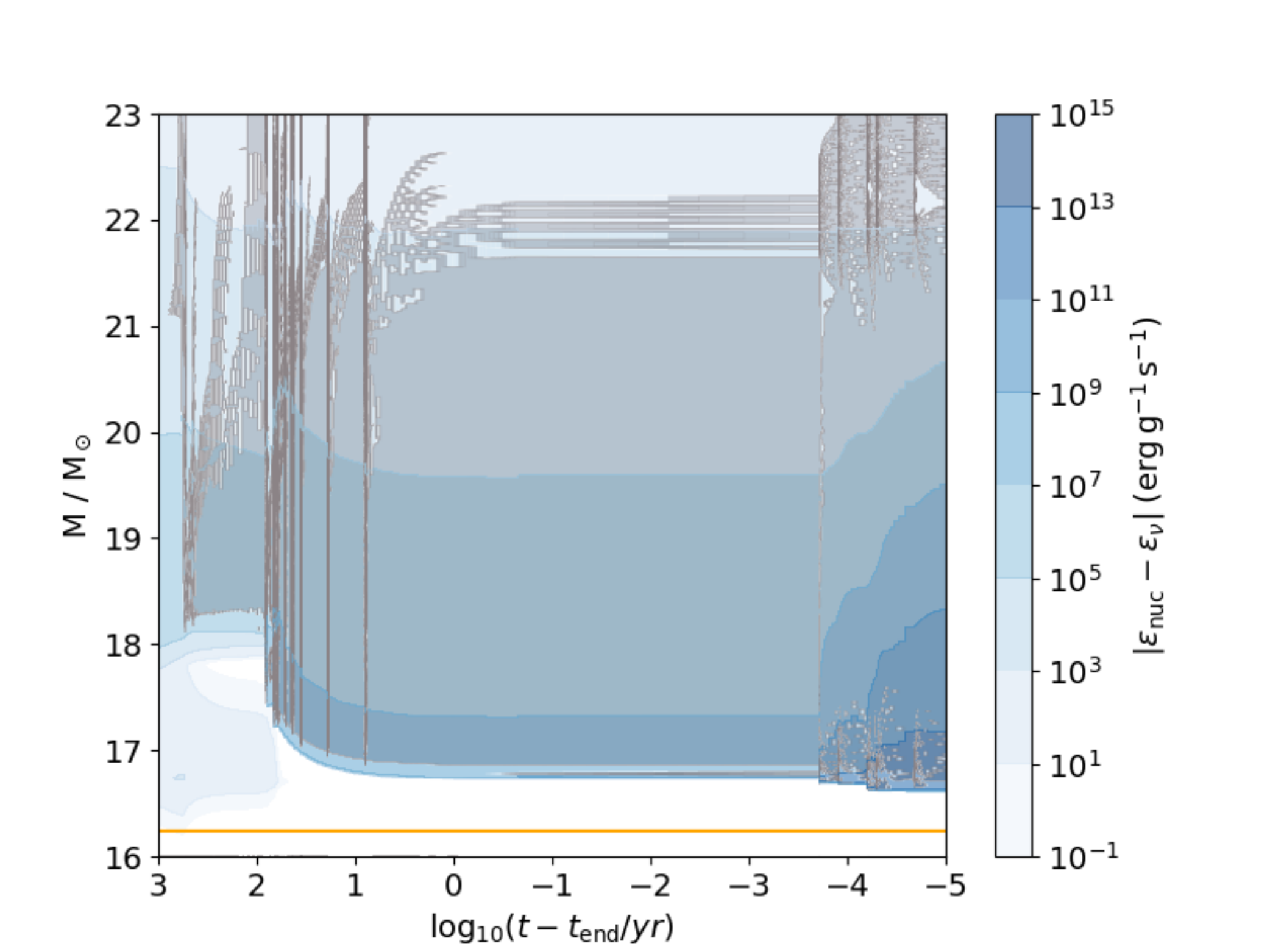}
    \caption{Zoom-in Kippenhahn diagram of H-He interaction in the \code{40schf-h} \Rshell\ case which includes the Schwarzschild criterion for convection and $f_{\mathrm{ov}}$ = 0.01. Colours are the same as in \Fig{40M_sch_f_kip}. The convective region is H-rich and the orange line shows the He-free core boundary.}
    \lFig{40Mschfzoom}
    \label{40Mschfzoom_fig}
\end{figure}

Also of note is the \code{40Mschf-h} case presented in Section \ref{corehburn}. In this simulation, unlike the \code{15Mschf-h} case and most others of this type, the H-rich material eventually penetrates all the way to the bottom of the radiative He shell releasing large amounts of energy, as indicated by the maximum $H$ number of 0.35 listed in \Tab{models} and $\epsilon_{\mathrm{nuc}}$ contours in \Fig{40Mschfzoom}. The H-shell boundary moves from a mass coordinate of $18.5\usp\Msun$ to $17\usp\Msun$ over a time span of  $502\usp \yr$ at which point the energy generation at the base of the H shell is $\sim 10^{9} \ergspergs$ and the $^{12}\mathrm{C}$ abundance is $X_{^{12}\mathrm{C}} = 10^{-4.6}$. Within another $35\usp\yr$, the H-shell boundary moves downward another $0.3\usp\Msun$ where the mass fraction of $^{12}\mathrm{C}$ abundance is now $X_{^{12}\mathrm{C}} = 0.31$ and the specific energy generation has increased to $\sim 10^{13} \ergspergs$. In this case, the temperature is $1.1\ee{8}\usp \unitstyle{K} $ at the maximum H-penetration point.  

\Tab{models} shows that $H$ can vary dramatically between simulations with \Rshell\ interactions. The primary difference in the simulations which have lower vs higher values of $H$ seems to be the abundance of $^{12}\mathrm{C}$ in the He shell the hydrogen is able to interact with. In the \code{15Mschf-f} model the mass fractions of $^{12}\mathrm{C}$ and protons where they meet are $X_{^{1}\mathrm{H}}  \approx 3\ee{-3}$ and $X_{^{12}\mathrm{C}}  \approx 2\ee{-6}$. If H only penetrates the upper layers of a radiative He shell, one does not expect the very high energy generation created when it reaches the bottom of the shell, where C is much more abundant, such as in the \code{40Mschf-h} case.  

The \Rshell\ interaction type bears similarity to dredge up events seen in AGB and super-AGB stars. \cite{2004ApJ...605..425H} showed that in models with assumed CBM, protons mixing inward and reacting with $^{12}\mathrm{C}$ causes the peak energy generation to be below the convective boundary, and ultimately leads to corrosive penetration into the radiative core. This same effect is shown in \Fig{15Mother} and occurs in all our \Rshell\ interactions. \citet[][and references therein]{2016MNRAS.455.3848J} show that not applying CBM at the lower boundary of the convection zone still results in a dredge-out episode. Furthermore, they demonstrated that CBM can lead to inward corrosive burning.

\subsection{Convective H shell and convective He shell}\label{C shell He}
The \code{40Mled} and \code{60Mled} simulations undergo an interaction between a convective H-burning shell and convective He-burning shell. In both, the H-He interaction occurs as carbon is exhausted in the core. This is the same mode of interaction reported in \cite{2018MNRAS.474L..37C} for a 45\Msun\ model. For this type of H-He interaction the \code{40Mled} model is used as a representative example and the overall evolution is shown in \Fig{40Mled}. After core He-burning, the inner regions of the star contract, igniting a convective He burning shell and the core begins burning C. As this happens, the H-burning shell is temporarily halted in its downward descent by the expansion of the He-burning shell below. 3000 yr after the He-burning shell becomes convective, a small portion of the convective He shell splits. This splitting can be seen in \Fig{40Mledzoom} at $\mathrm{M}/\Msun \sim 14.2$. A similar small convection zone is located at the base of the H-shell. It is uncertain at this time whether such splitting of convection zones is simply an artefact of 1D mixing assumptions or if similar features may be also be found in 3D hydrodynamic simulations. Approximately three weeks after this split forms, H-burning shell material begins mixing into the radiative layer separating the two convection zones, or intershell region, as indicated by the increase in nuclear energy generation at the base of the H shell along with an increase in the abundances of $^1$H, $^{13}$C and $^{13}$N in the intershell, and later at the top of the He shell. 
In this case, H is initially ingested into the intershell region and $7\usp yr$ later, He-burning products mix upward into the H shell as well, indicated by an increase in CNO nucleosynthetic products seen in the top right panel of \Fig{40Mledother}. Only $12.6\usp\unitstyle{min}$ after the event begins, the energy generation rate from CNO reactions has increased to  $\log \epsilon_{\mathrm{CNO}} =11.6$. At later times protons have to some extent mixed through much of the He shell.  

After the initial interaction, small entropy steps within the intershell mix, leaving a new, ever descending barrier for the remainder of the simulation, which is 5.6 \unitstyle{hr}. In the bottom right panel of \Fig{40Mledother}, this entropy barrier can be seen in addition to the fact that this barrier is not preventing mixing between the H and He-rich layers. As a reminder to the reader, the \code{40Mled} model does not include the effects of CBM. The peak specific energy generation in the H-He interaction is $\log \epsilon_{\mathrm{CNO}}  = 13.25$.

\begin{figure}
	\includegraphics[width=\columnwidth]{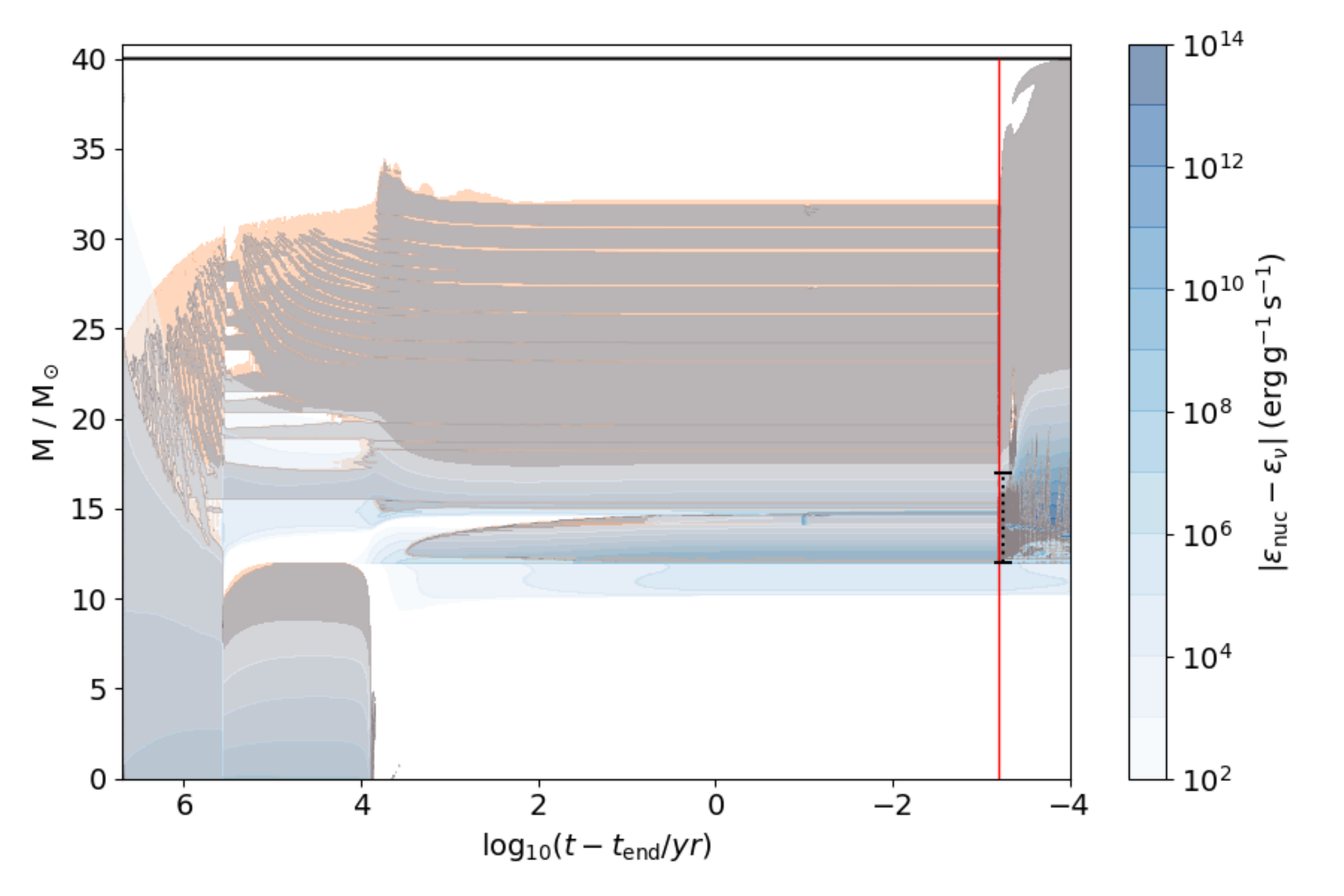}
    \caption{Kippenhahn diagram of the whole calculated evolution of the \code{40Mled} \Cshell\ case, using the Ledoux criterion for convection and $\alpha_{\mathrm{semi}} = 0.5$. Colours are the same as in \Fig{40M_sch_f_kip} and peach areas show where semiconvection has been applied. The red line indicates the beginning of the H-He interaction and the black dotted line indicates where the profiles in \Fig{40Mledother} are taken.}
    \lFig{40Mled}
    \label{40Mled_fig}
\end{figure}

\begin{figure}
	\includegraphics[width=\columnwidth]{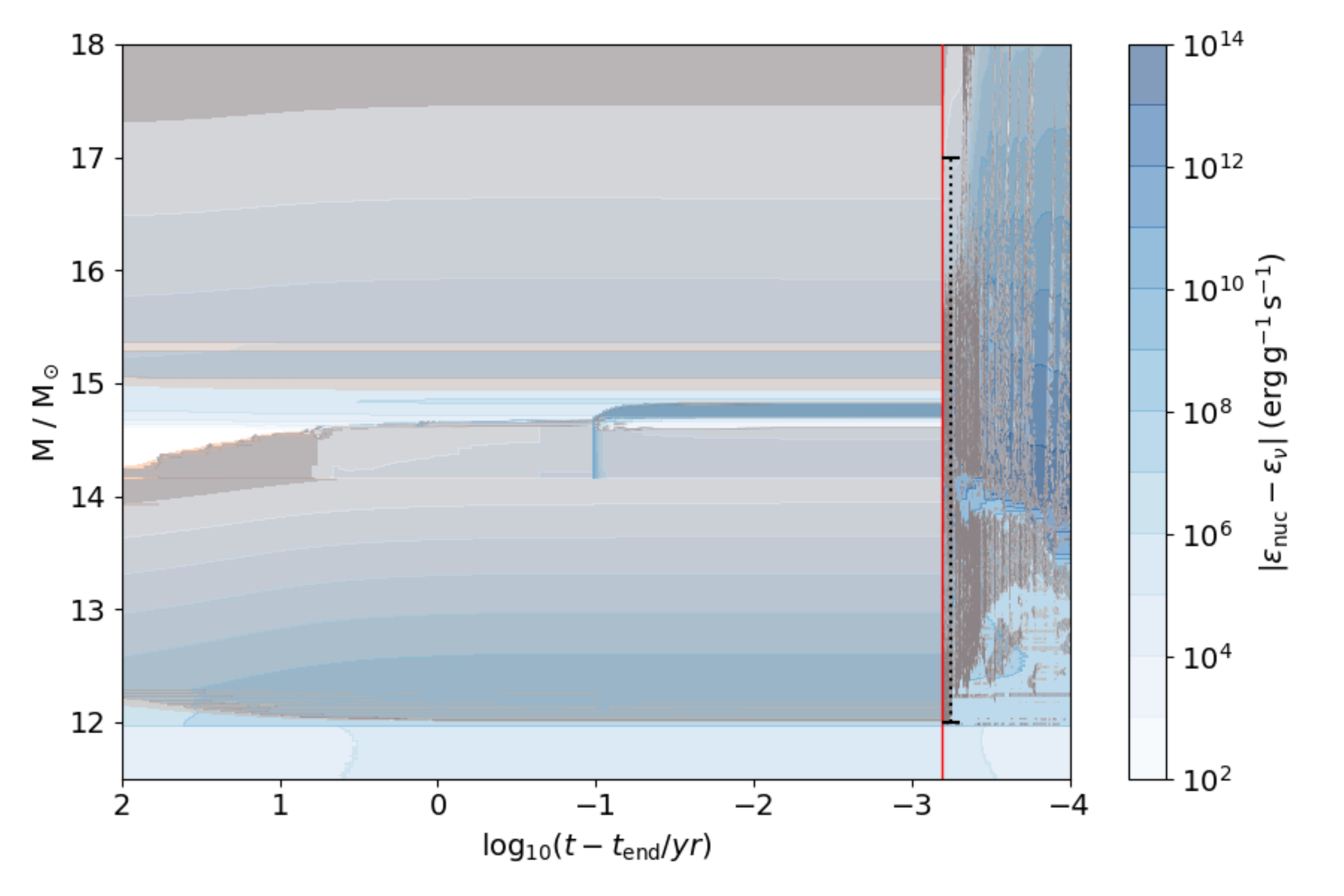}
    \caption{Zoom-in of \Fig{40Mled}. Colours are the same as in \Fig{40M_sch_f_kip} and peach areas show where semiconvection has been applied. The red line indicates the beginning of the H-He interaction and the black dotted line indicates where the profiles in \Fig{40Mledother} are taken.}
    \lFig{40Mledzoom}
    \label{40Mledzoom_fig}
\end{figure}

\begin{figure*}
	\includegraphics[width=\textwidth]{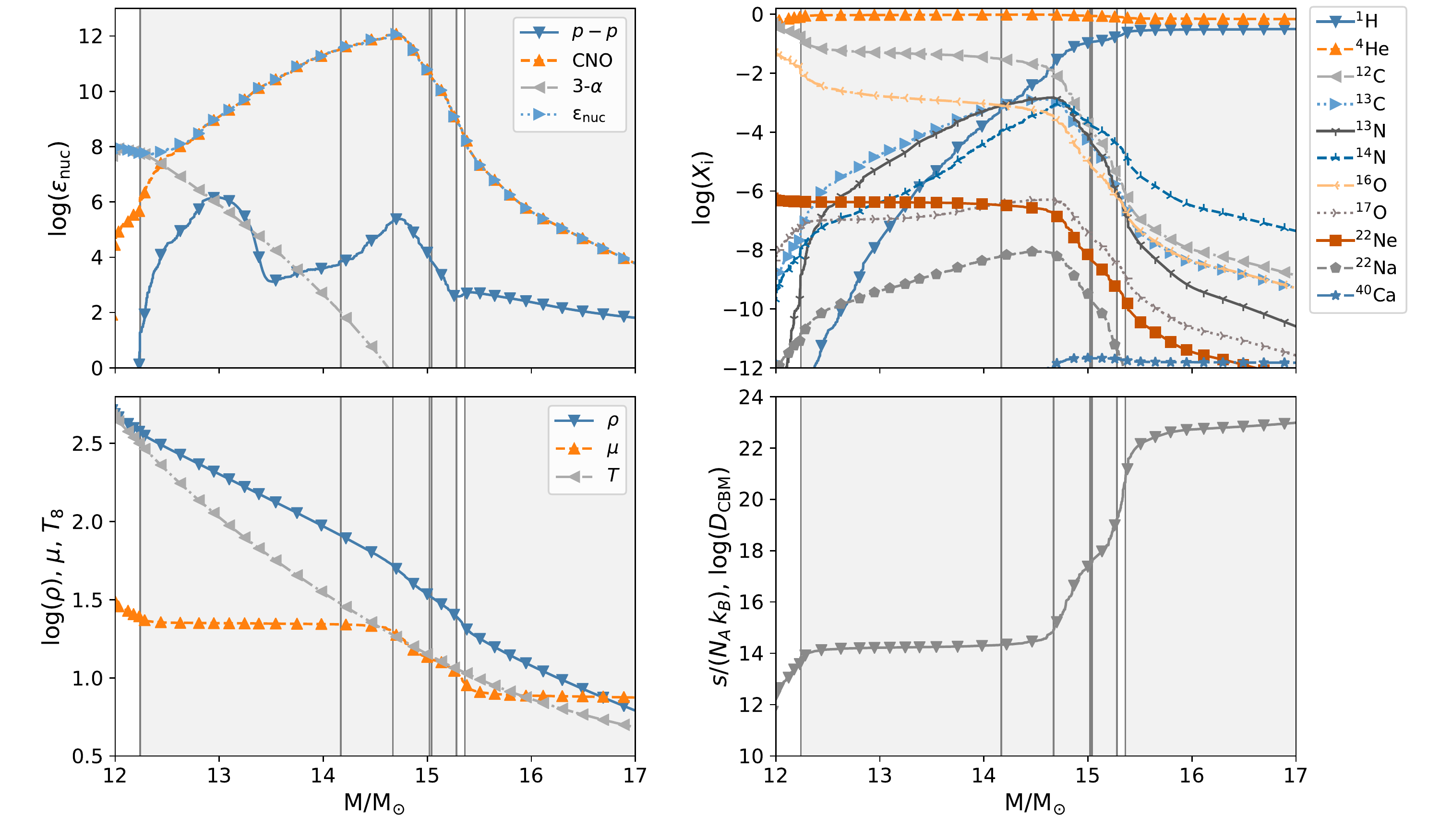}
    \caption{Profiles for the \Cshell\ interaction \code{40Mled} model with the Ledoux criterion for convection and semiconvecton included. \textbf{Top left}: Total specific energy generation and that from CNO, p-p and tri-$\alpha$ reaction groups. \textbf{Top right}: Mass fractions of several abundant species. \textbf{Bottom left}: Temperature, mean molecular weight and density profiles. \textbf{Bottom right}: Specific entropy. Profiles shown are $40 \usp\unitstyle{min}$ after the H-He interaction begins in the \code{40Mled} model, times for which are indicated in \Fig{40Mled} as black dashed and solid red lines, respectively. Grey areas show regions unstable to convection. See Appendix \ref{appendix} for profiles of diffusion coefficents and similar profile plot before interaction occurs}.
    \lFig{40Mledother}
\end{figure*}

This interaction type is among the more energetic which can be seen in the maximum $H$ numbers reported in \Tab{models}, 0.22 and 0.28. The high energy generation is the consequence of protons being convectively mixed into the He burning shell, rather than a slow corrosive type burn as seen in \Rshell\ models where H-rich material encroaches into a radiative He-burning layer.

The presence of convection in the He shell leads to a higher C
abundance at the top of the shell, as compared to a radiative He
shell, and causes the downward mixing of protons to higher
temperatures. The \iprn\ occurs when H is convectively mixed into a C-rich environment on a timescale comparable to the decay of $^{13}\mathrm{N}$. It begins with the $^{12}\mathrm{C}(p,\gamma)^{13}\mathrm{N}$ reaction. $^{13}\mathrm{N}$ then $\beta^{+}$ decays to $^{13}\mathrm{C}$ while being transported to higher temperatures where the $^{13}\mathrm{C}(\alpha ,n)^{16}\mathrm{O}$ is strongly
activated. In these simulations we treat the emergence of the
neutron source consistently with the mixing processes. We also
include n-capture reactions at least one isotope away from the
valley of stability up to V. Since n-capture reactions for heavier
and trans-iron element are missing the resulting neutron density is an upper limit. However, the main source of uncertainty comes from the treatment of convective mixing and the stellar evolution
response in 1D. \cite{2019MNRAS.488.4258D} demonstrated that the entrainment rate of H into the He shell will strongly influence the neutron density for the \iprn\ in the case of rapidly accreting white dwarfs, therefore we report a number only indicative of what the neutron density might be.

\subsection{Convective H shell and radiative layer above He core}\label{R core He}
This interaction type occurs in 3/26 simulations and is characterized by H-rich material entering the radiative layer above the convective He-burning core in the later stages of core He-burning. While similar, this differs from \Ccore\ interactions in that it occurs at a later evolutionary phase and the H-rich material never enters the convective core. Similar to the \Rshell\ models, it occurs only in models with CBM. The \code{140Mledf-h} simulation, which used the Ledoux criterion for convective stability, semiconvection and our higher value of CBM is described for this case and the overall evolution is shown in \Fig{140Mledf}. During core He-burning, a small H-burning convection zone, as described in Section \ref{corehburn}, descends downward over the course of $1.8\ee{5}\usp\yr$. In this time, the core contracts by about $0.2\Rsun$ and the base of the H shell has an relevant CBM region of $0.1\--0.6\usp\Msun$. The H shell begins to descend into the previously H-free region, with protons mixing into the radiative layer atop the He core. The convective core moves down in tandem with the shell for $4.5\ee{5}\usp\yr$. Shortly after, core convection effectively ceases, which can be seen in that last hour of the simulation in \Fig{140Mledf}. The ceasing of core convection is seen in all three \Rcore\ interactions. 

Just before core convection stops, the CBM region at the base of the H shell begins to extend over the region which was previously the convective core. Then, convection in the H shell brings He-burning ashes upward. The introduction of large amounts of C and O into the H shell from the He shell pushes the CNO cycle out of equilibrium. In this simulation, energy generation from CNO increases up to $\log \epsilon_{\mathrm{CNO}}  = 12.5$. We do not see a return to CNO equilibrium for the duration of the simulation. A similar chain of events takes place in the \code{40Mschf-l} simulation although in the \code{40ledf-h} simulation, the H shell never descends low enough to bring up material from the previously convective core. In the \code{140Mledf-h} simulation, the upward mixing of He-core material triggers Ne-Na and Mg-Al cycles in the mixing region. Evidence for this can be seen in the top left panel of \Fig{140Mother} showing nuclear energy generation from reactions involving Na, Ne and Mg and the top right panel shows this mixing and nucleosynthesis $54\usp\min$ after the event has begun. $^{22}\mathrm{Na}$ has a half-life of $2\usp\yr$ and is being produced in-situ in the convective mixing region. Based on elemental ratios for the region at the end of the simulation, this nucleosynthesis does not remove the odd-even effect as is observed in some of the most iron-poor stars (e.g. HE1327-2326 and HE0107-5240) but may somewhat lessen it. It is also interesting to note that if such an interaction were to occur and the core were to remain convective, the \iprn\ may be effectively triggered, although 3D hydrodynamic investigations are needed to confirm this. In \cite{2015A&A...580A..32M, 2017A&A...605A..63C} Ne-Na and Mg-Al cycles are discussed in massive Pop III and low-Z stars as being a result of rotational mixing processes. Our simulations show that convection-induced mixing processes may lead to similar nucleosynthesis. Overall, further investigation is required to properly understand the full nucleosynthetic outcome in such scenarios.

\begin{figure}
	\includegraphics[width=\columnwidth]{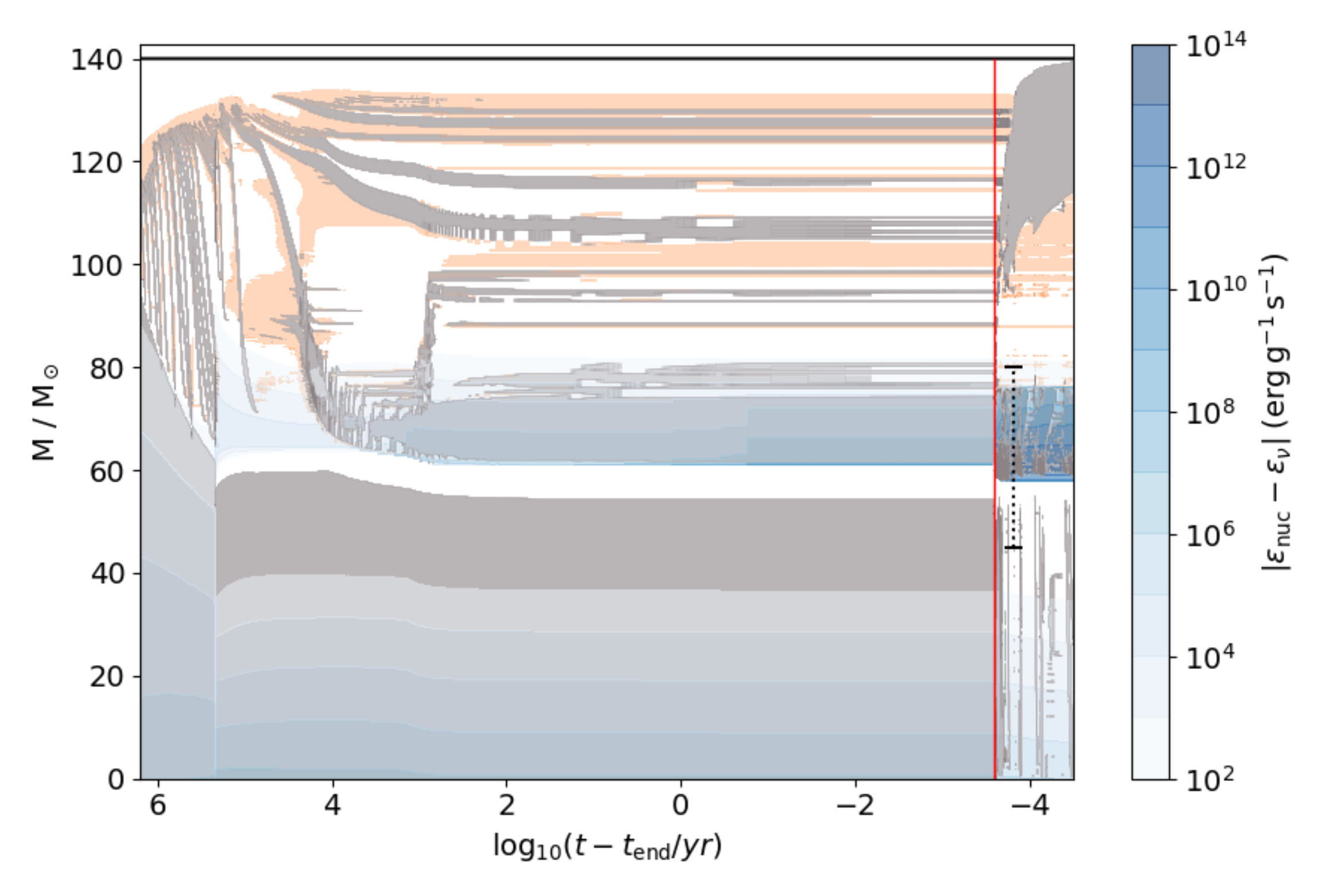}
    \caption{Kippenhahn diagram of the $140\usp\Msun$ \Rcore\ model with the Schwarzschild criterion for convection and $f_{\mathrm{ov}}$ = 0.01 (\code{140Mledf-h}) case. Colours are the same as in \Fig{40M_sch_f_kip} and peach areas show where semiconvection has been applied. The red line indicates the beginning of the H-He interaction and the black dotted line indicates where the profiles in \Fig{40Mledother} are taken.}
    \lFig{140Mledf}
    \label{140Mledf_fig}
\end{figure}

\begin{figure*}
	\includegraphics[width=\textwidth]{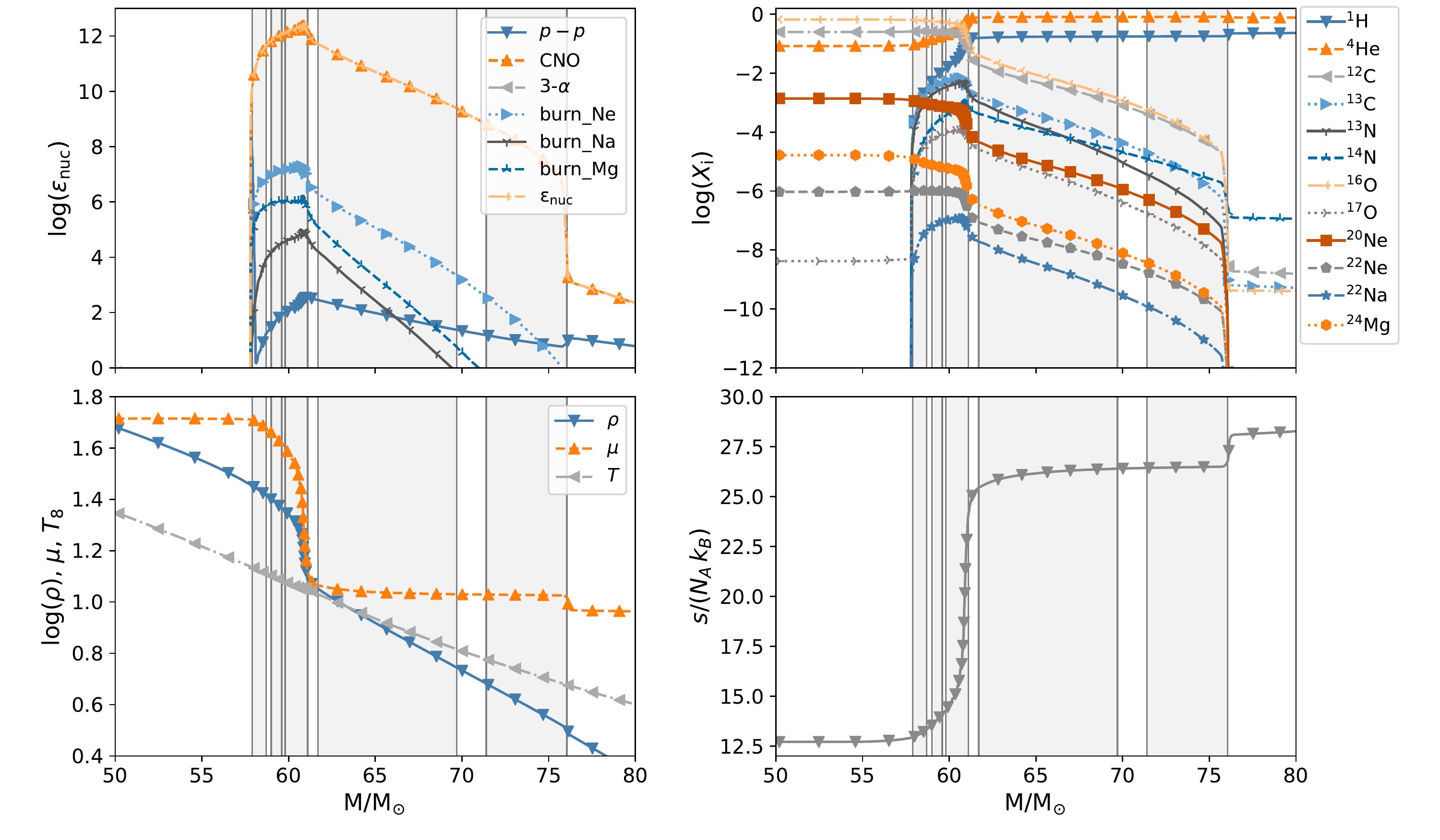}
    \caption{Profiles for the \Rcore\ He interaction \code{140Mledf-h} \Rcore\ case, with the Ledoux criterion for convection, $\alpha_{\mathrm{semi}} =0.5$, and $f_{\mathrm{ov}} =0.01$. \textbf{Top left}: Total specific energy generation and that from CNO, p-p and tri-$\alpha$, Na, Mg and Ne burning reaction groups. \textbf{Top right}: Mass fractions of several abundant species. \textbf{Bottom left}: Temperature, mean molecular weight and density profiles. \textbf{Bottom right}: Specific entropy. Profiles taken $54 \usp\unitstyle{min}$ after the H-He interaction begins, times for which are indicated in \Fig{140Mledf} as black dashed and solid red lines, respectively. Grey areas show regions unstable to convection. See Appendix \ref{appendix} for profiles of diffusion coefficents}.
    \lFig{140Mother}
\end{figure*}

\subsection{Convective H shell and convective He core}\label{C core He}
There is also the case of H-He interactions where the H-rich material enters the convective He-burning core, which comes in second for frequency in 5/26 cases. All of these occur in simulations with $M \geq 60\usp\Msun$ using the Schwarzchild criterion for convection including CBM. We describe the \code{80Mschf-h} case, though all cases displaying interaction between a convective H-burning shell and convective He-burning core begin in a similar fashion. At the beginning core He-burning, the H-burning layer above the core shows little convection, as seen in \Fig{80schkip}, although it is near the threshold for convective instability. Core He-burning proceeds for $1.5\ee{3}\usp\yr$, the H-shell becomes convective and then within $40\usp\unitstyle{yr}$, H-burning material enters into the H-free core. In this case, and in others of this class, there is no distinct entropy barrier formed (\Fig{80mother}) as the H-rich material enters into the H-free region. Rather, the existing barrier at the top of the He core is gradually smoothed, with small entropy steps forming in the previously radiative intershell. 10 minutes before the interaction, the He-core has a relevant CBM region of about $0.67 \usp \Msun$. In this case, CBM does not play a role in mixing at the base of the H shell as the Lagrangian motion of the H shell is much faster than CBM can effectively mix.

\begin{figure}
	\includegraphics[width=\columnwidth]{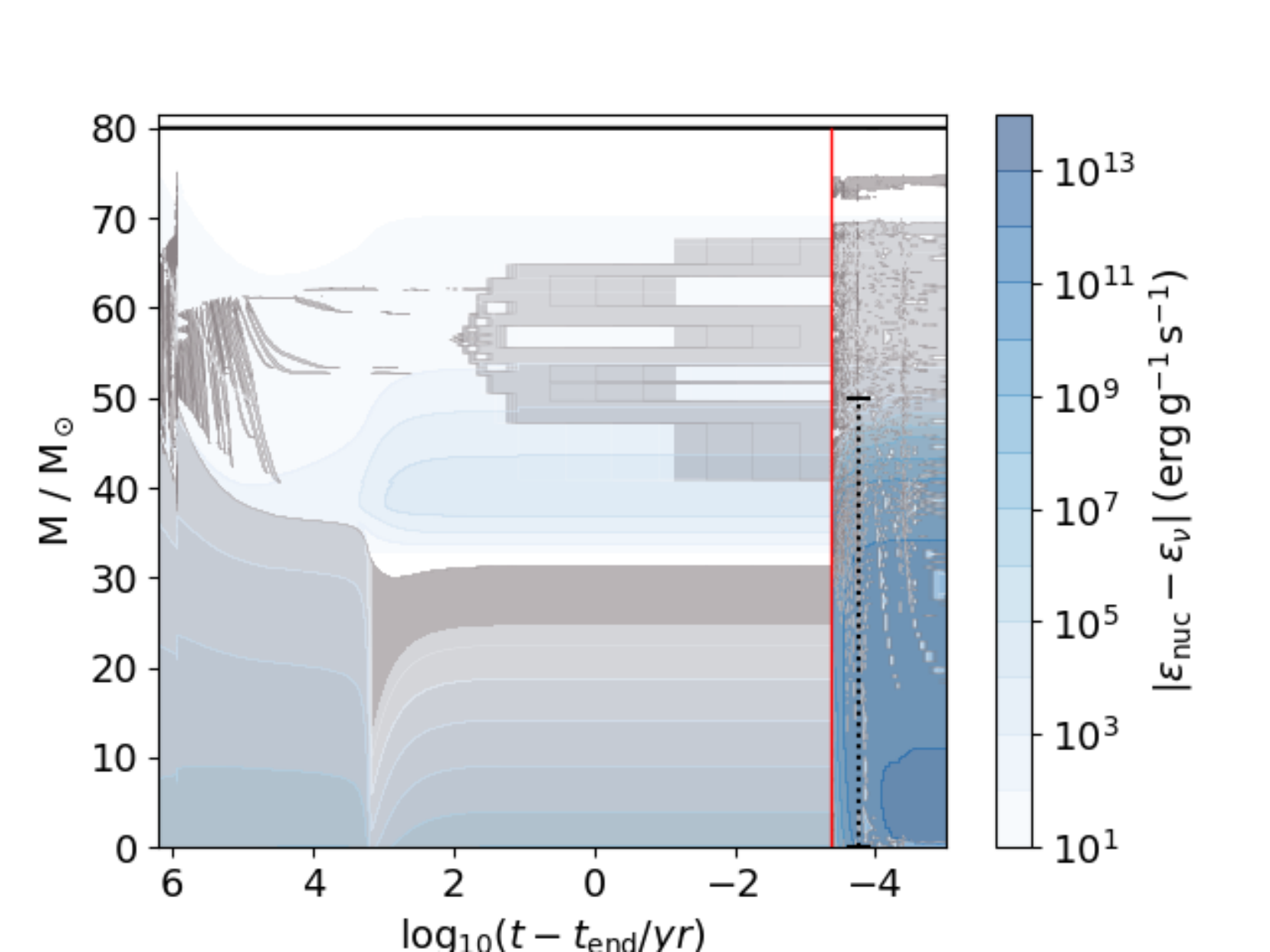}
    \caption{Kippenhahn diagram of the $80\usp\Msun$ \Ccore\ model with the Schwarzschild criterion for convection and $f_{\mathrm{ov}}$ = 0.01 (\code{80Mschf-h}). The red line indicates the beginning of the H-He interaction and the black dotted line indicates where the profiles in \Fig{80mother} are taken. Colours are the same as in \Fig{40M_sch_f_kip}.  }
    \lFig{80schkip}
\end{figure}

\begin{figure*}
	\includegraphics[width=\textwidth]{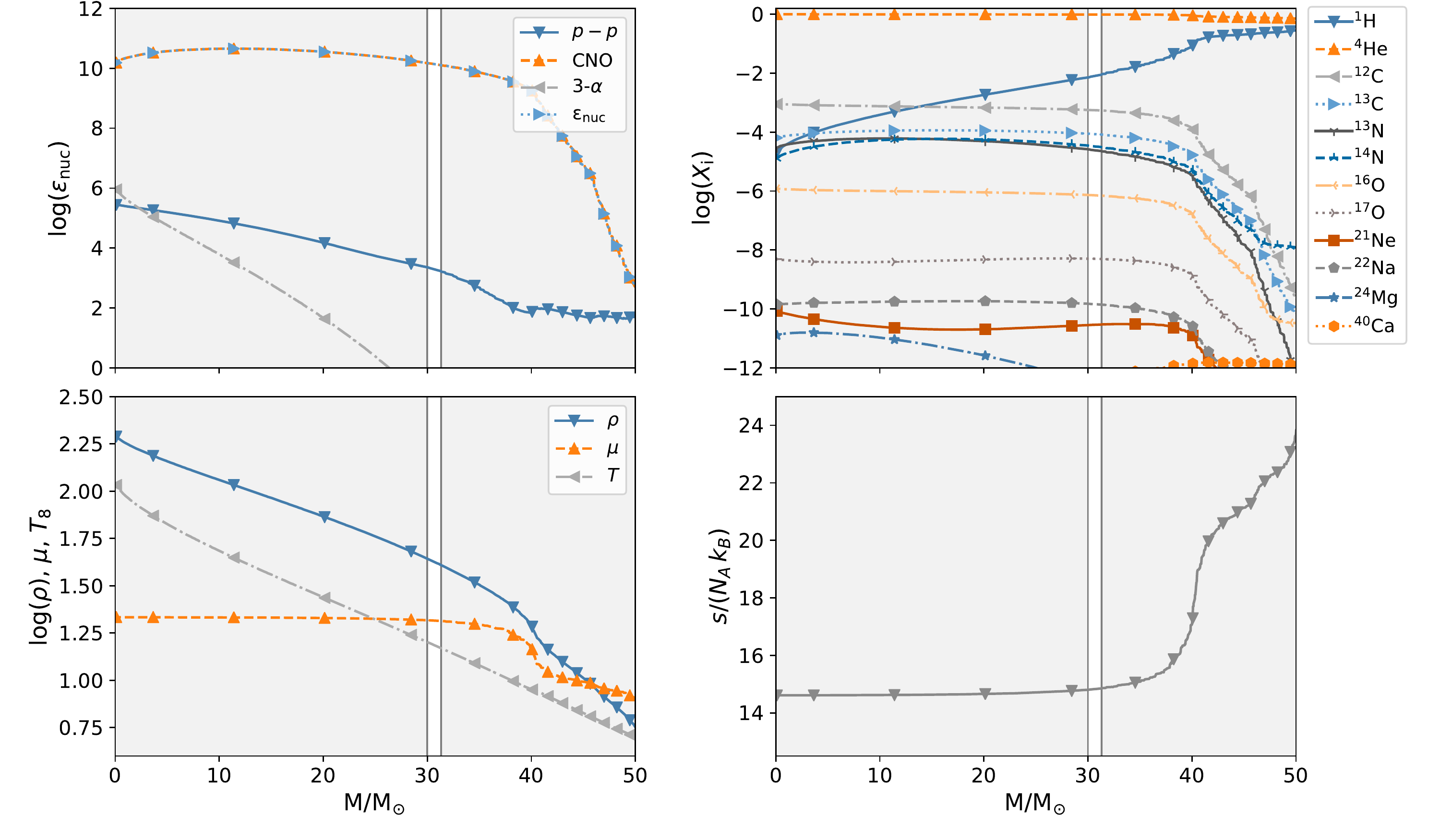}
    \caption{Profiles for the \code{80Mschf-h} \Ccore\ case, with the Schwarzschild criterion for convection and $f_{\mathrm{ov}} =0.01$. \textbf{Top left}: Total specific energy generation and that from CNO, p-p and tri-$\alpha$ reaction groups. \textbf{Top right}: Mass fractions of several abundant species. \textbf{Bottom left}: Temperature, mean molecular weight and density profiles. \textbf{Bottom right}: Specific entropy. Quantities are shown 1.75 \unitstyle{hrs} after the H-He interaction begins, the time and mass coordinates for which is indicated in \Fig{80schkip} as a dashed black line. Grey areas show regions unstable to convection. See Appendix \ref{appendix} for profiles of diffusion coefficents}.
    \lFig{80mother}
\end{figure*}

All interactions of this type occur at the beginning of core He-burning meaning that there is little C in the core at this point, from $X_{^{12}\mathrm{C}} \sim 10^{-6} - 10^{-3}$---levels too low for the \iprn\ to play an important role in such interactions. In \Tab{models} it can be seen that the \code{60Msch-} and \code {140Msch-f} cases show no change in mass coordinate for the interaction. For these three simulations we do not use $X_{\mathrm{H}} = 10^{-4}$ as the H-free boundary because the H shell and He core merge completely and the value of $\Delta$ $\mathrm{M}/\Msun = 0$ in \Tab{models} would result from our measurement choices. In the  \code{60Msch-} and \code {140Msch-f} models, the H abundance never drops below $10^{-4}$ in the core. Therefore for these simulations, $X_{\mathrm{H}} = 10^{-3}$ was used to compute  $\Delta$ $\mathrm{M}/\Msun$.

The weak \spr \,  has been reported in the literature to happen in massive stars of both solar and low metallicity as the result of rotational mixing carrying $^{14}\mathrm{N}$ from the H shell into the He core where $\alpha$ captures will transform it to $^{22}\mathrm{Ne}$ thus providing for the $^{22}\mathrm{Ne}(\alpha,n)^{25}\mathrm{Mg}$ neutron source \citep{2008ApJ...687L..95P, 2008IAUS..250..217H}. This is controlled by relatively slow mixing and burning---quite different from what occurs in our C core interactions. In the \code{80Mschf-h} case we see no evidence of significant upward mixing of C and O before the mixing event as is reported in \cite{2016MNRAS.456.1803F}. The \code{80Mschf-h} simulation has a peak neutron density after the interaction of $\log{N_\mathrm{n}} \sim 11$. It may be that lower $^{12}\mathrm{C}$ mass fractions can lead to \spr\ neutron densities with C as a seed as opposed to Fe, but further studies must be conducted to confirm this. 

\subsection{No H-He interaction}\label{None}
The four simulations that do not undergo any kind of H-He interaction are \code{15Mled}, \code{80Mled}, \code{140Mled} simulations, and the \code{60Mledf-l} simulations. The first three of these cases have semiconvection with no CBM included, and the \code{60Mledf-l} simulation has semiconvection plus the lower of our chosen CBM values. Additionally, they all use the Ledoux criterion for convection which includes the stabilizing effect of the gradient in $\mu$. Clearly this is not always sufficient in preventing a H-He interaction, because the \code{40-} and \code{60Mled}, and all other than the $60\usp\Msun$ \code{-led-l} simulations do experience some kind of H-He interaction. These cases are run until the point of Fe-core infall, except for the \code{140Mled}. As mentioned in Section \ref{shellhburn}, the \code{140Mled} simulation becomes pair unstable. \Fig{15Mled} shows an example of the \code{15Mled} simulation which has no H-He interaction during its evolution.

Zero-metallicity stars of $\sim 140\usp\Msun$ likely end their lives as pair instability supernova \citep{2003ApJ...591..288H}. For the \code{140Mled} stellar model, all analysis in this work is reported for times before the onset of core O-burning, when the star becomes pair unstable. The adiabatic index, $\Gamma_1$ falls below 4/3 in the C-shell leading up to core O-burning, in an off-centre manner with central temperatures rising to over $3\ee{9} \usp\unitstyle{K}$. Additionally, the condition that He core masses $\gtrapprox 60\usp\Msun$ should explode as pair instability supernova \citep[and references therein]{RevModPhys.74.1015,2012ARA&A..50..107L} agrees with our result as this model has a He core mass of $\approx64\usp\Msun$. From the onset of the instability (dynamic phase) we do not include subsequent evolution in our analysis. For the \code{140Mled} model, \Tab{models} only displays H-shell temperatures to the onset of the instability. 

\begin{figure}
	\includegraphics[width=\columnwidth]{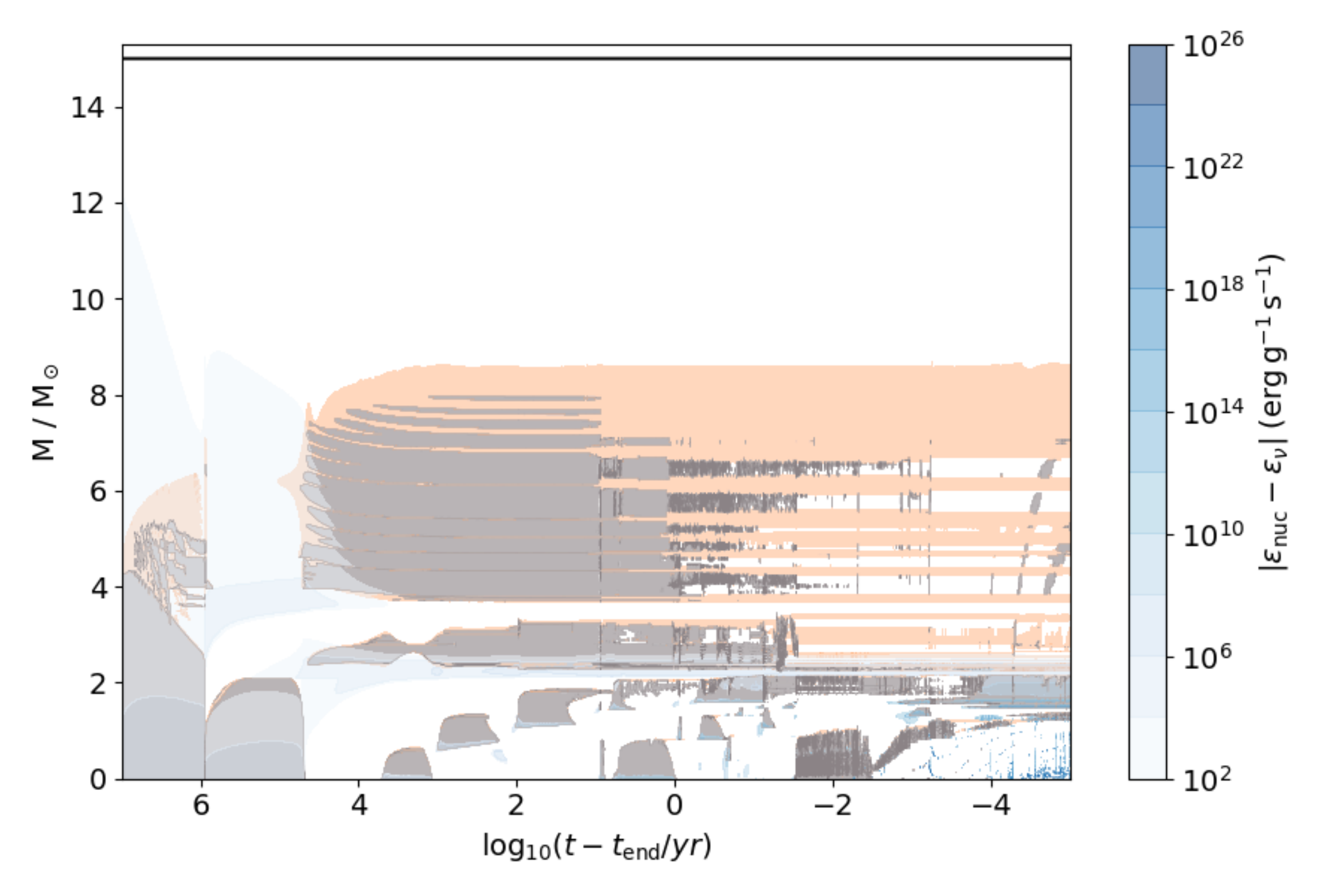}
    \caption{Kippenhahn diagram of the $15\usp\Msun$ model with the Ledoux criterion for convection, and semiconvective mixing included (\code{15Mled}) where no H-He interaction takes place during the evolution. Colours are the same as in \Fig{40M_sch_f_kip} and peach areas show where semiconvection has been applied. }
    \lFig{15Mled}
\end{figure}

\subsection{C and N ratios}\label{CN}
As described in the preceding subsections, H-He interactions have a variety of effects on the nucleosynthesis that takes place within the H-He region, ranging from moderate to large increase in CNO cycle activity resulting either from downward mixing of protons such that the $^{12}\mathrm{C}(p,\gamma)^{13}\mathrm{N}$ is triggered effectively, the upward mixing of CNO catalysts boosting the overall CNO activity, or a combination thereof.

In \cite{2018MNRAS.474L..37C} we suggest that with a high $H$ number, it may be possible to eject material from the H-He interaction region prior to the star's death. In this section we use the same assumption as \cite{2018MNRAS.474L..37C} to demonstrate CNO ratios from H-He interactions. Despite the different physical assumptions, our integration procedure is the same as one would employ when assuming a faint supernova with no mixing or nucleosynthetic contribution from the shock (c.f., \cite{2014ApJ...794...40T, 2017A&A...605A..63C}). This is the same calculation used in Section \ref{hotcno}, using Equation \ref{eq2}. Here, due to the uncertainties in the duration of a H-He interaction, we measure the abundances at different times throughout the interaction, not just at the end of the simulation. 

\Fig{CNO} shows the range of $^{12}\mathrm{C}/^{13}\mathrm{C}$ and [C/N] values in simulations explained in Section \ref{h-he} with the addition of the \code{60Mledf-l} and \code{80Mled} simulations. The latter two are shown as examples of the C and N ratios that would result from a faint supernova with no mixing or explosive nucleosynthesis in simulations with no H-He interaction for comparison. We also show the high $H$ number \code{40Mschf-h} and \code{80Mledf-h} simulations. 
 
 During a H-He interaction, timesteps can vary by many orders of magnitude and each interaction has a unique duration in physical time. Therefore, to sample the C and N abundances during a H-He interaction, we use a linear spacing in time steps, of 10 points from just before the H-He interaction begins to when there is $\sim5\usp\min$ of simulation time left, as we do not expect any significant changes in the abundances on timescales smaller. We calculate CNO abundances for two different cutoff assumptions. The green and blue points in \Fig{CNO} are calculated from innermost point where the mass fraction of protons in the H-He layer is $10^{-4}$, at the typical region of maximum energy generation. pink and yellow points have been sampled somewhat deeper, where the proton mass fraction is $10^{-8}$. We assume the material taking part in the interaction becomes unbound. The point where the simulations end in this work does not necessarily represent when the event would stop in reality. This is when the 1D calculations are no longer able to continue and suggests the transition to dynamic, 3D behaviour. 
 
 Therefore, \Fig{CNO} displays the range of possible C and N ratios during such an event. The \code{60Mledf-l} and \code{80Mled} simulations are shown as examples of simulations that do not experience a H-He interaction and we apply the same procedure to calculate abundances but from the end of core He-burning. For the \code{60Mledf-l} simulation, no point can match the observations because the H envelope material remains in CNO equilibrium. The \code{80Mled} simulation experiences hot CNO cycling in the final stages of it's evolution and initially has similar CNO equilibrium ratios as many of the other models. Later, the [C/N] ratio increases by almost $2\usp\unitstyle{dex}$ due to partial hot CNO cycling. The  $^{12}\mathrm{C}/^{13}\mathrm{C}$ ratios make H-burning conditions in this model incompatible with the observations as hot CNO burning generally lowers this value. This may be partially alleviated by the inclusion of He-shell material ($^{12}\mathrm{C}$), though a full investigation is beyond the scope of this work. We compare our simulations to observed CEMP-no stars from the literature with measured $^{12}\mathrm{C}/^{13}\mathrm{C}$ ratios. While well aware of the shortcomings in modelling the time-evolution of H-He interactions in 1D, the CNO ratios shown illustrate that H-He interactions in massive Pop III stars can display a range of values for  $^{12}\mathrm{C}/^{13}\mathrm{C}$ and [C/N], which results in the isotopic and elemental abundance ratios found in stars believed to be the second generation of stars. We do not include any contribution from the interstellar medium because pristine Big Bang abundances would have virtually no effect on C and N isotopic or elemental ratios. 
 
 In the bottom left of \Fig{CNO}, there is a clustering of points
 representing the CN equilibrium ratios for conditions within massive Pop III stars. During a H-He interaction, when H-burning material moves deeper into the He layer, the overall [C/N] ratio rises as the mass coordinate of the H-rich front lowers, as there is simply a much larger mass fraction of C as compared to the H-shell CNO abundances. Similarly, the $^{12}\mathrm{C}/^{13}\mathrm{C}$ generally increases as the mass coordinate of the H-rich front lower as there is a high $^{12}\mathrm{C}$ abundance in the He shell. Both of these general trends are complicated by the simultaneous mixing and nucleosynthesis taking place and neither evolve monotonically. This is exemplified by the \code{40Mled} data. A black dotted line is included to \Fig{CNO} to guide the eye. Following blue to green points from the `early' point at $^{12}\mathrm{C}/^{13}\mathrm{C} \sim 0.5$ and [C/N] $\sim -2.1$. Shortly after that, $^{12}\mathrm{C}$ increases as the H-burning front moves downward. This leads to large values for both of these ratios. Then, both ratios begin to fall creating a loop-like pattern governed by first the $^{12}\mathrm{C}(p,\gamma)$ reaction, leading to a lower $^{12}\mathrm{C}/^{13}\mathrm{C}$ ratio. This is followed by $^{13}\mathrm{C}(p,\gamma)$ which leads to a lower [C/N] ratio. In the future, these ratios may help to constrain the duration, depth and timing of H-He interactions, which as of now, are uncertain parameters.
 
In some of the cases presented in this work, such as the \code{40Mled} simulation, protons mix down into the He-rich region where there is a relatively large abundance of C. In this situation, the  $^{12}\mathrm{C}(p,\gamma)^{13}\mathrm{N}$ takes place, $^{13}\mathrm{N}$ is transported downward to higher temperatures where it decays in the presence of $\alpha$ particles. Here CNO cycling does not take place as the timescale for the $^{13}\mathrm{C}(\alpha,n)^{16}\mathrm{O}$ reaction is many orders of magnitude shorter than that of $^{13}\mathrm{C}(p,\gamma)$, where there are very few protons at the bottom of the He-shell. In other cases presented here, the variety in CN ratios relates to the mixing of protons with large amounts of $^{12}\mathrm{C}$ and $^{16}\mathrm{O}$, which boosts CNO activity. In many of our simulations, the CN ratios are never able to return to equilibrium values and the points in \Fig{CNO} show how this mixing and burning can evolve in time. 

\begin{figure}
	\includegraphics[width=\columnwidth]{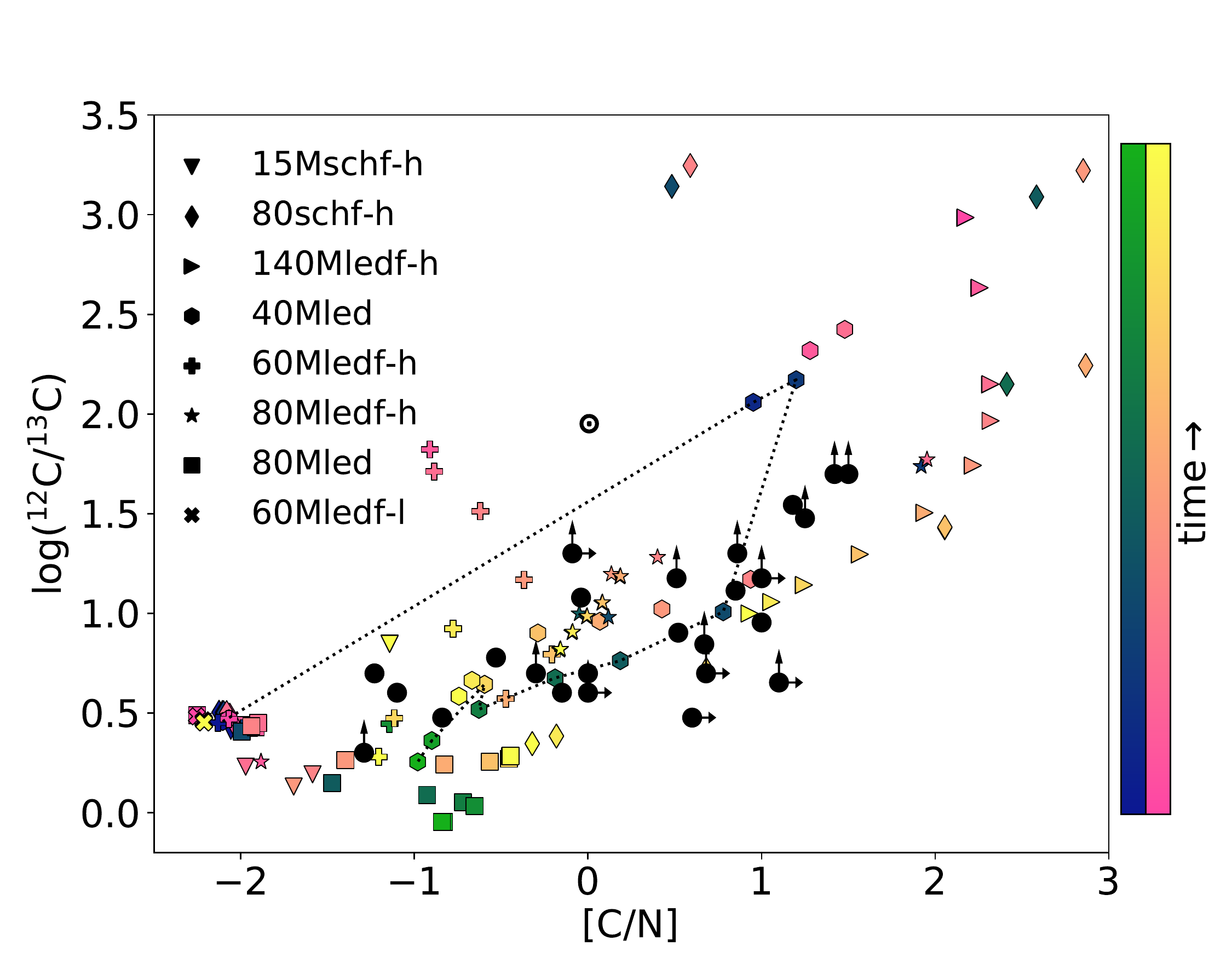}
    \caption{C isotopic and CN elemental ratios from observed CEMP-no stars are shown in black. Data are taken from \citealt{2007AJ....133.1193B},  \citealt{2005A&A...434.1117P},\citealt{2013ApJ...762...28N}, \citealt{2010ApJ...722L.104N}, \citealt{2013ApJ...762...26Y}, \citealt{2014ApJ...784..158R}, \citealt{2015ApJ...807..173H} and \citealt{2010A&A...509A..93M}. Blue and green symbols show ratios in the H-He region or H-shell for models which do not experience a H-He interaction. Pink and yellow symbols sample abundances deeper within the region. The choice of data points is described in the text of Section \ref{CN} and the calculation for each abundance is given in Equation \ref{eq2}. The black dotted illustrates one of two paths taken by the \code{40Mled} case. Simulation data point colours change as an indication of time with pink and blue being earlier in a H-He interaction and yellow and green being later. Note that in some simulations, points at the same time for different depths can largely overlap. The solar ratios are indicated by the $\odot$ symbol}.
   \lFig{CNO}
\end{figure}

\section{Discussion}\label{discussion}
The structural properties of Pop III stars seem to lend themselves to H-He interactions. The cause of H-He interactions has been associated with relatively small temperature gradients \citep{2012ApJS..199...38L} and entropy differences \citep{2010ApJ...724..341H} between H and He layers in Pop III stars. Shallow temperature gradients are the result of the higher temperatures found in H-burning. All other state variables constant, a higher H-burning temperature would lead to a higher entropy in the H-burning shell, yet the smaller gradient in temperature leads to a less stiff boundary. The latter is also the case for the shallow density gradients found between H and He layers in Pop III stars. The overall smaller entropy differences between the two layers are primarily due to the higher densities in the H envelope, which are orders of magnitude higher than in post-MS solar metallicity massive stars. Pop III stars also have a much lower envelope opacity than solar metallicity stars. All of this is the result of an initial zero metallicity which results in a hotter, more compact star. 

\cite{2017A&A...605A..63C} explore the nucleosynthesis which may occur in massive stars of metallicity $\mathrm{Z} =10^{-5}$. The authors highlight production of H-burning material in the form of CNO and elements with Z = 10$^{-13}$ amplified by the upward mixing of $^{12}$C and other isotopes from the He shell into the H-burning shell due to rotational mixing. For some models, an ad hoc enhanced shear diffusion coefficient is used in order to initiate H-He interactions at specific times. They then relate [C/N] and $^{12}\mathrm{C}$/$^{13}\mathrm{C}$ to observations. In general concurrence with our results presented in Section \ref{CN}, they found that H-He interactions occurring in massive Pop III stellar models reproduce the CN ratios found in CEMP-no stars. 

\cite{2017A&A...605A..63C} suggest that the final $^{12}\mathrm{C}$/$^{13}\mathrm{C}$ ratio depends on their selected timing of the mixing event relative to the end of the stars life. This is true if one were to assume that the star does not eject material with these signatures prior to Fe-core infall, thus terminating the CNO cycle. In addition, we also find that the downward mixing of protons into the He-shell also terminates the CNO cycle by cutting off the supply of protons. Both of these effects lead to a variety of CN ratios in our simulations. Additionally, it appears that secular mixing process such as rotation \citep{2017A&A...605A..63C, 2015A&A...580A..32M}, or semiconvection (this work) cannot simultaneously reproduce many of $^{12}\mathrm{C}$/$^{13}\mathrm{C}$ and [C/N] ratios observed in CEMP-no stars.

H-He interactions are less likely in stars of increasing metallicity than in their zero-metallicity counterparts and due to both the differences in structure such as a higher opacity and entropy barrier between the H and He shells, and H-burning nucleosynthesis. Overall, H-He interactions become less likely with increasing metallicity. Despite this, they have been reported in simulations at low-Z and have been reported in super-AGB stars by \cite{2016MNRAS.455.3848J} and massive stars by \cite{2018MNRAS.480..538R}.

H-He interactions occur \textit{not infrequently} in simulations of massive Pop III stars by \cite{2012ApJS..199...38L}. This is confirmed by our findings for any given set of parameters. They also note that differences have been found in results for H-He interactions from various authors but do not specify what they are. While we cannot know at this time how much of these differences are due to different physics or implementations of stellar physics and numerical methods, our results show that there are multiple modes of H-He interactions, and more than one kind can be present using the same code and macro-physics assumptions. In an attempt to explain the CEMP-no star HE 0107-5240 the same authors report that H-He interactions lead to C, N, and Na relative to Mg in ratios consistent with the mixing of protons into He shell \citep{2003ApJ...594L.123L}---although the nucleosynthetic process responsible was not described. 

\cite{2018ApJ...865..120B} present yields of \ipr\ nucleosynthesis from parametrised post-processing calculations using $20-30 \usp\Msun$ stellar evolution models. They focus on what in this work we call \Cshell\ interactions---where H is \textit{ingested} into a convective He shell below. They only consider stars of $>20\usp\Msun$  because, as they state, those less massive will not undergo any ingestion because He shells in these stars never expand to the base of the H-rich layers. They also state that at masses $>30\usp\Msun$ proton ingestion events will not happen because H has already been depleted in the shell. We have found that other types of H-He interactions can occur without a convective He shell, but a convective He region allows for the \iprn\ to occur. In our \code{140Mledf-h} and \code{40Mschf-h} simulations, shown in Figs. \ref{40Mschfzoom_fig} and \ref{140Mledf_fig}, where H-rich material corrosively burns and descends into a radiative He core and -shell, respectively, the material is able to descend far enough into the He shell by the end of the simulation that protons and $^{12}\mathrm{C}$ mix at mass fractions of $\sim10^{-2}$ and $\approx 0.3$, respectively. This can lead to high $H$ numbers although our simulations do not indicate that the \iprn\ can take place in such scenarios.
 
 In \cite{2018ApJ...865..120B} all shell interactions were initiated by hand using ad-hoc diffusion during nucleosynthesis post-processing. As the authors state, this is done as they wanted to avoid a splitting of the convection zone. In 1D simulations the  \iprn\ may be prevented or suppressed by the formation of an entropy barrier \citep{2018MNRAS.474L..37C, 2011ApJ...727...89H}. At this time it is not clear how such a barrier would evolve in 3D. Additionally, the stellar evolution simulations in \cite{2018ApJ...865..120B} were post-processed and do not include the effects of the energy generation due to the H-He interaction on the 1D model, which is orders of magnitude greater than in quiescent H-shell burning. This result has been shown in the present work, and in \cite{2018MNRAS.474L..37C} for a similar scenario where luminosities can reach $\log(L/L_{\odot}) \sim 13$. Inspection of the Kippenhahn diagrams shown in Figs. \ref{40Mled_fig} and \ref{40Mledzoom_fig} of the \code{40M-led} simulation, shows that the energy generation from nuclear reactions has a significant effect on the structure of the star for the remainder of the simulation. 

Mixing of H-rich material into He cores and shells has been reported by \cite{2012A&A...542A.113Y} for both rotating and non-rotating stellar models of  $20-500\usp\Msun$. As in the work presented here, calculations are terminated for H-He shell interactions shortly after the onset. They found that models with shell interaction can produce up to 1000x more $^{14}\mathrm{N}$ in the H shell as compared to models that do not have a shell interaction due to upward mixing of CNO catalyst but still less than those with chemically homogeneous evolution as a result of rotation. They do not describe the nucleosynthesis in the He shell as they did not have mixing and burning equations coupled, and in convective-reactive scenarios, $\mathrm{Da}$ goes to unity. Furthermore, they observe convective H shells in several models that extend downward and interact with the He-burning core. They describe two modes of interaction for H-He core interactions---both of which begin with a descending envelope as we have described in Sec. \ref{C core He} and Sec \ref{R core He}. The difference in the two relates to the primary direction of material during the interaction. In some models, once the H shell descends, material moves downward into the He core and in others, material moves primarily upwards into the H shell boosting the CNO cycle as core convection has turned off. These reports appear very similar to our \Rcore\ and \Ccore\ simulations. 

\cite{2007A&A...461..571H} found that in simulations of very low metallicity rotating massive stars, primairy N production was enhanced by H-He interactions leading to the upward mixing of CNO catalyst from the He-burning core into the H shell. This enhanced N production was found to better reproduce the observations of CNO abundance ratios evolution in metal-poor halo stars \citep{2006A&A...449L..27C}. \cite{2015ApJ...808L..43P} argued that H-He interactions followed by SN explosions at low metallicity should be at least partly responsible for the solar $^{14}\mathrm{N}/^{15}\mathrm{N}$ value and may play a greater role at low-Z. We have found that a boost in CNO can occur during the H-He interactions presented in this work, for example in the \code{15Mschf} and \code{140Mledf} models. Total yields are dependent on the final fate of the star but using the same assumptions made in \ref{CN} we can derive an estimate of the N yield. For the \code{15Mschf} model, at the final timestep the total integrated N yield is $4\ee{-7}\usp\Msun$. For both the \code{40Mled} and \code{140Mledf} models we find a total N yield of $\approx 0.04\usp\Msun$. The \code{80Mschf} model gives $2.5\ee{-4}\usp\Msun$. This value can be compared with the \code{80Mled} model presented in \ref{hotcno}, which does not encounter a H-He interaction. The \code{80Mled} model has a much lower final N yield of $6.7\ee{-7}\usp\Msun$. Two major caveats are worth noting for our estimations. First, it should be noted that we do not calculate beyond the H-He interaction and it is currently unclear what the long-term evolution of such models would be. If the simulations were to evolve longer, this may increase N production in some models by allowing CNO time to equilibrate. Second, we assume the complete ejection of the envelope, although if a GOSH or similar instability were to occur in some cases without a supernova explosion, it may be that only a fraction of this material is ejected. In this case, our estimates represent upper limits. Given the uncertainties in the final fates of these stars, these numbers should be taken as illustrative estimates as opposed to definitive predictions.

The results from 1D stellar evolution during events of this nature violate the assumptions of MLT \citep{2016MNRAS.455.3848J}. One of the issues is the fact that the timescale for mixing and burning becomes comparable and at the same time, large amounts of energy are released that can modify the convective morphology of the layer. At this point the details of how these interactions will evolve are unclear and modelling assumptions must be made. For the case of low-metallicity, low-mass stars \cite{2009PASA...26..139C} do not alter the result of stellar evolution simulations that a split in the convection zone
forms. \cite{2018ApJ...865..120B} for the case of massive low or zero metallicity assume in the post-processing of H ingestion into the He-shell convection zone a continuous mixing and burning without feedback into the stellar structure.  \cite{2011ApJ...727...89H} also allowed mixing to continue past the time where the split of the convection zone would have occurred, guided by elemental abundance observations of the post-AGB star Sakurai's object.

Using 3D hydrodynamical simulations, \cite{2014ApJ...792L...3H} found that the ingestion and burning of H into the He-shell convection zone lead early-on to the formation of an entropy shelf in spherically-averaged profiles in the upper half of the He-rich convection zone. 3D hydrodynamical simulations of a low-Z AGB star by \cite{2011ApJ...742..121S} do not find a split in the convection zone, although they caution this could be because their simulations were not evolved beyond $4\usp\unitstyle{hr}$ star time. It is not clear currently what the long term evolution of a H-He interaction is and how they may compare in different stellar environments. Despite this, 3D simulations suggest that some aspects of these events are recovered in 1D simulations, such as the location of H-burning and high energy generation which leads to $\mathrm{Da} \sim 1$ \citep{2014ApJ...792L...3H}. In 3D hydrodynamic simulations, this combustion leads to a dynamic, non-spherically symmetric response within the star, that would violate assumptions of MLT, and at least temporarily, hydrostatic equilibrium. To properly understand if and how these events unfold and answer questions about e.g. how the entropy barrier evolves in time for massive Pop III stars, these events must be simulated in 3D.

\section{Conclusion} \label{conclusion}
We have investigated the occurrence, variation and types of H-He interactions in massive Pop III stars with a variety of masses and mixing assumptions. This study demonstrates that using several reasonable assumptions regarding macrophysical mixing processes, the unique structural properties of Pop III stars, which result from their initial composition, lend themselves to H-He interactions in 1D stellar evolution models. We have found that there are four distinct modes of interaction depending on the evolutionary stage and whether or not protons mix with convective He-burning material. Only loose trends can be found with respect to either mass or mixing prescription. For example, interactions involving radiative He layers occur only when using CBM and core interactions become more frequent over $40\usp\Msun$. H-He interactions correlate with contractions at the onset of, or during core He burning, at the beginning of core C burning or at the beginning of core O burning. 

Energy production in these events depends primarily on the amounts of $^{12}\mathrm{C}$ and protons in the He and H shell respectively and is dominated by CNO reactions in all simulations. In \Rshell\ and \Ccore\ simulations, the $H$ number tends to be lower.  Later in core He burning, \Rcore\ interactions can take place which are generally more energetic as higher amounts of C and O mix into the previously radiative intershell region above the core. This mixing can also bring Ne and Mg from the core triggering Ne-Na and Mg-Al cycles, which to our knowledge, has only been reported in works including rotation. Lastly, \Cshell\ interactions, which have been identified as a possible \ipr\ site, are typically highly energetic, resulting from more advanced convective He-shell burning.

We have also found that as a H-He interaction unfolds, $^{12}\mathrm{C}$/$^{13}\mathrm{C}$ and [C/N] ratios can take a wide range of values owing to both downward mixing of protons and upward mixing of He-ashes. We have shown that the $^{12}\mathrm{C}$/$^{13}\mathrm{C}$ and [C/N] ratios found in our simulations are consistent with observed CEMP-no stars, stars believed to carry the chemical signatures of the first stars. 

We have explored multiple frequently-used mixing prescriptions to better understand the parameter space in which these events occur and how they occur. As such, the frequency or behaviour of H-He interactions presented in this paper should not be considered as representative of the frequency or behaviour in the first stars. Given the 3D hydrodynamic nature of these events is not yet well understood, this work should serve as a roadmap for future 3D hydrodynamic simulations to investigate H-He interactions which may have taken place in the first generation of stars.

\section*{Acknowledgements}
We would like to thank Raphael Hirschi for his helpful comments while refereeing this paper. OC would like to thank Pavel Denissenkov, Sam Jones, and Etienne Kaiser for comments on drafts of this paper and Michael Wiescher for valuable input on nuclear physics aspects. FH acknowledges funding through an NSERC Discovery Grant. Simulations were carried out on the Compute Canada Arbutus cloud computer as part of a computing resource allocation to Falk Herwig. This work benefited from support by the National Science Foundation under Grant No. PHY-1430152 (JINA Center for the Evolution of the Elements). This work was made possible by the open source \code{MESA} stellar evolution code.

\section*{Data Availability}
The data underlying this article will be shared on reasonable request to the corresponding author.



\bibliographystyle{mnras}
\bibliography{hhe.bib} 



\appendix
\section{Appendix A}
\label{appendix}

\begin{figure*}
	\includegraphics[width=\textwidth]{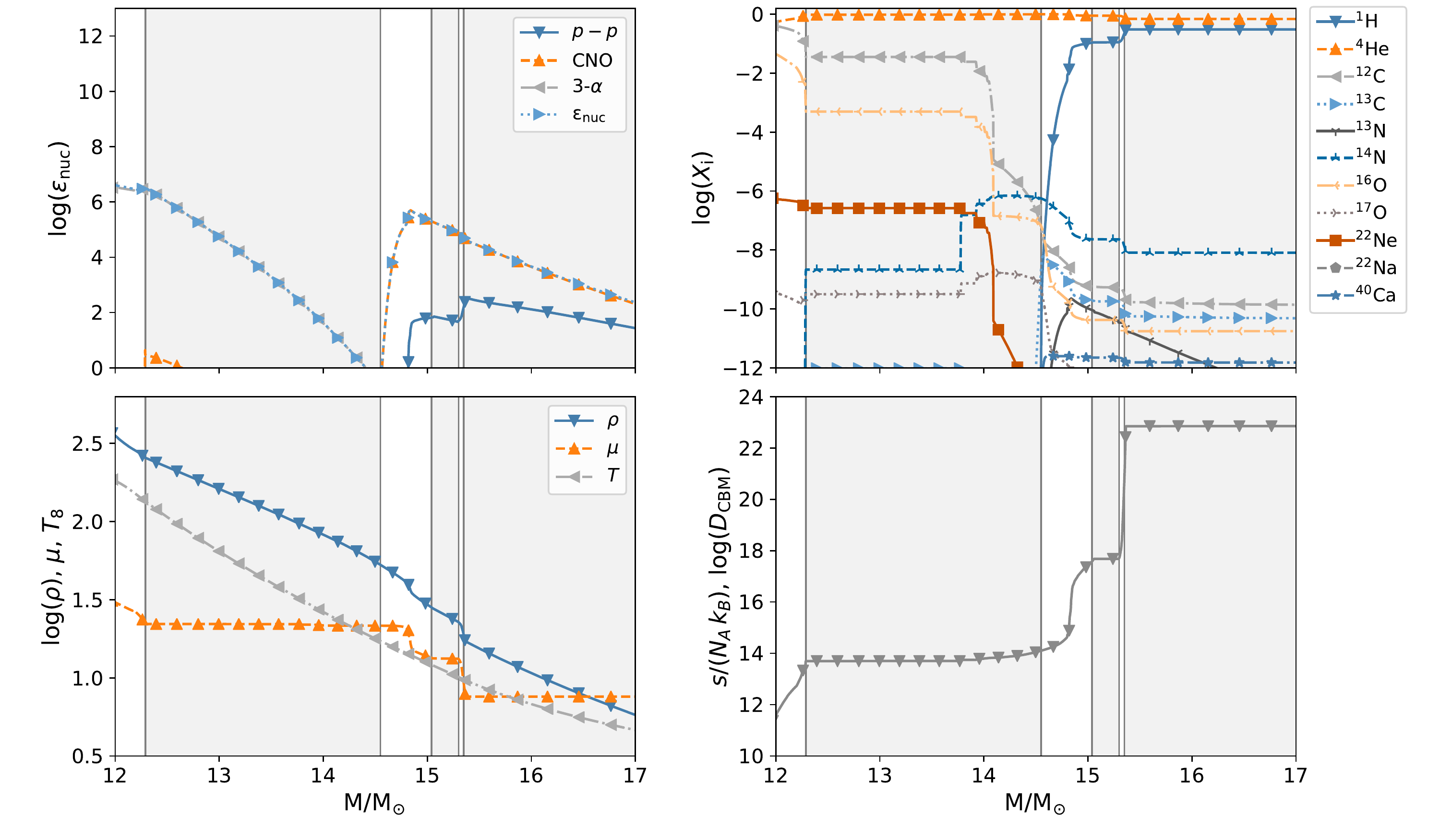}
    \caption{Profiles for the \Cshell\ interaction \code{40Mled} model with the Ledoux criterion for convection and semiconvecton included. \textbf{Top left}: Total specific energy generation and that from CNO, p-p and tri-$\alpha$ reaction groups. \textbf{Top right}: Mass fractions of several abundant species. \textbf{Bottom left}: Temperature, mean molecular weight and density profiles. \textbf{Bottom right}: Specific entropy. Profiles taken $\approx 9\usp\unitstyle{yr}$ before the H-He interaction begins, at the same time as the top panel of \Fig{CCshellD}. Grey areas show regions unstable to convection.}
    \lFig{40Mledother_before}
\end{figure*}

\begin{figure}
	\includegraphics[width=\columnwidth]{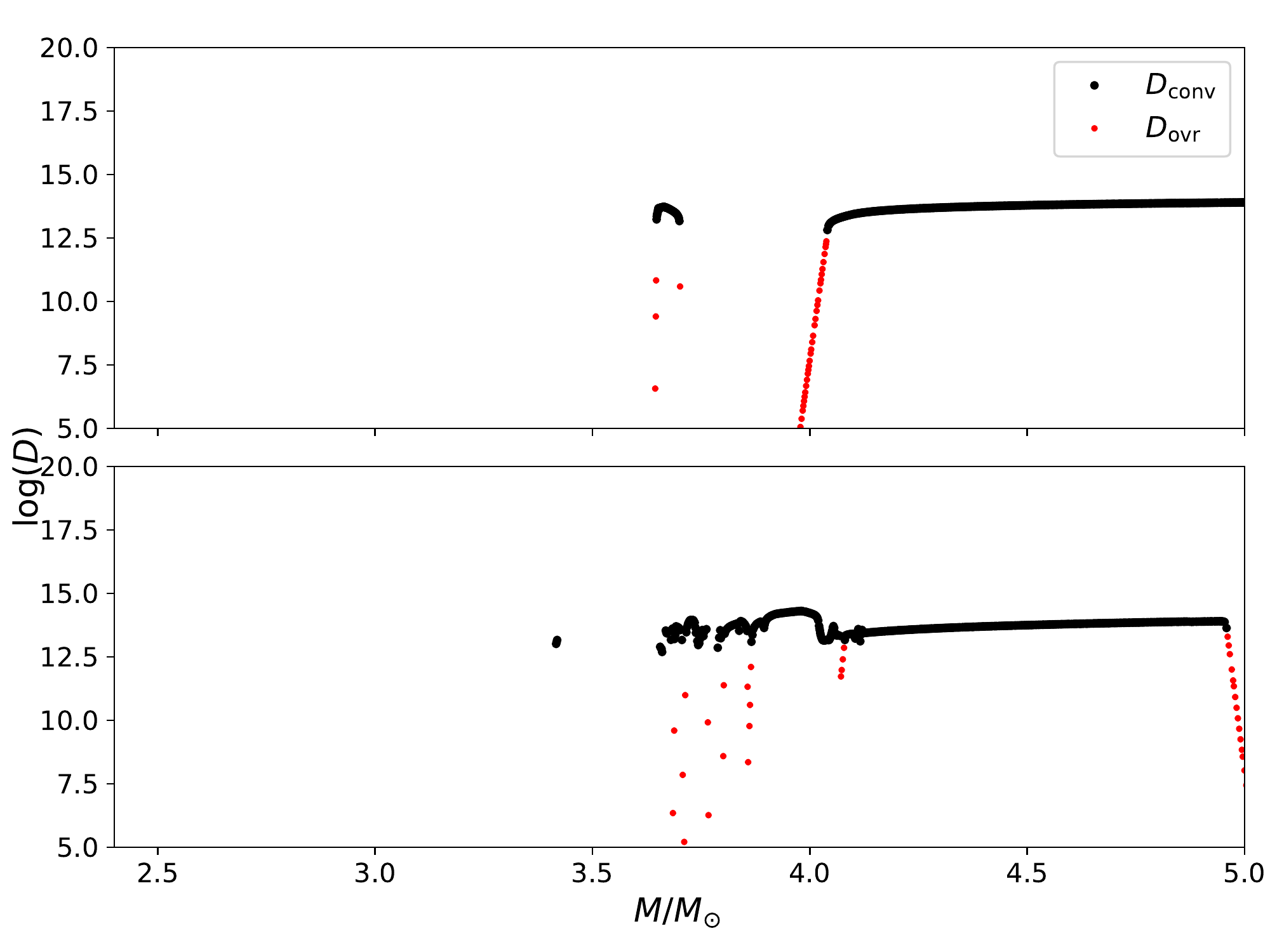}
    \caption{\textbf{Top}: Diffusion coefficients for convection and CBM in the \code{15Mschf-h} model $277\usp\unitstyle{yr}$ before the H-He interaction begins. \textbf{Bottom}: Same quantities at the time shown in \Fig{15Mother}. }
   \lFig{CRshellD}
\end{figure}

\begin{figure}
	\includegraphics[width=\columnwidth]{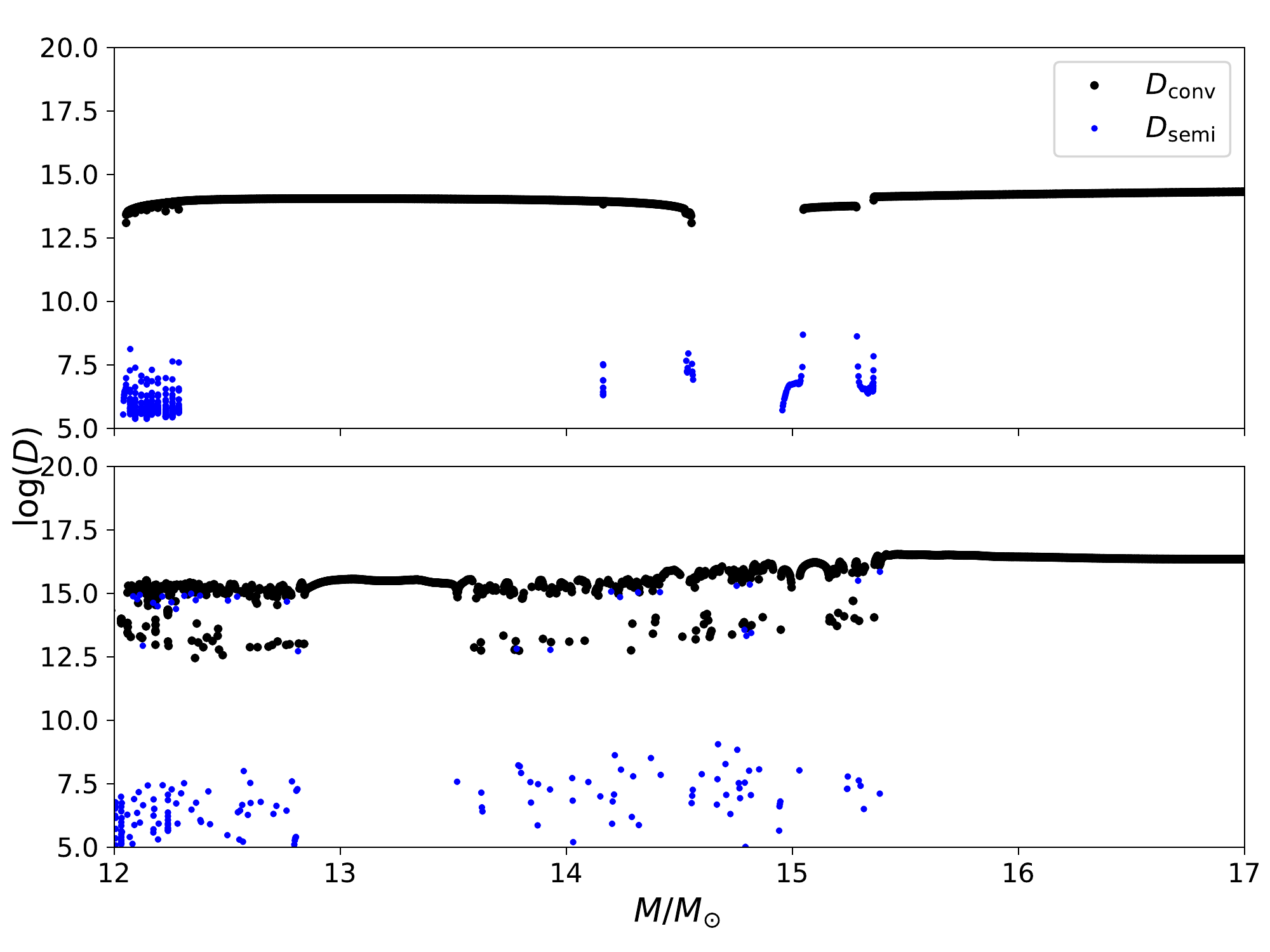}
    \caption{\textbf{Top}: Diffusion coefficients for convection and semiconvection in the \code{40Mled} model $9 \usp\unitstyle{yr}$ before the H-He interaction begins. \textbf{Bottom}: Same quantities at the time shown in \Fig{40Mledother}. }
   \lFig{CCshellD}
\end{figure}

\begin{figure}
	\includegraphics[width=\columnwidth]{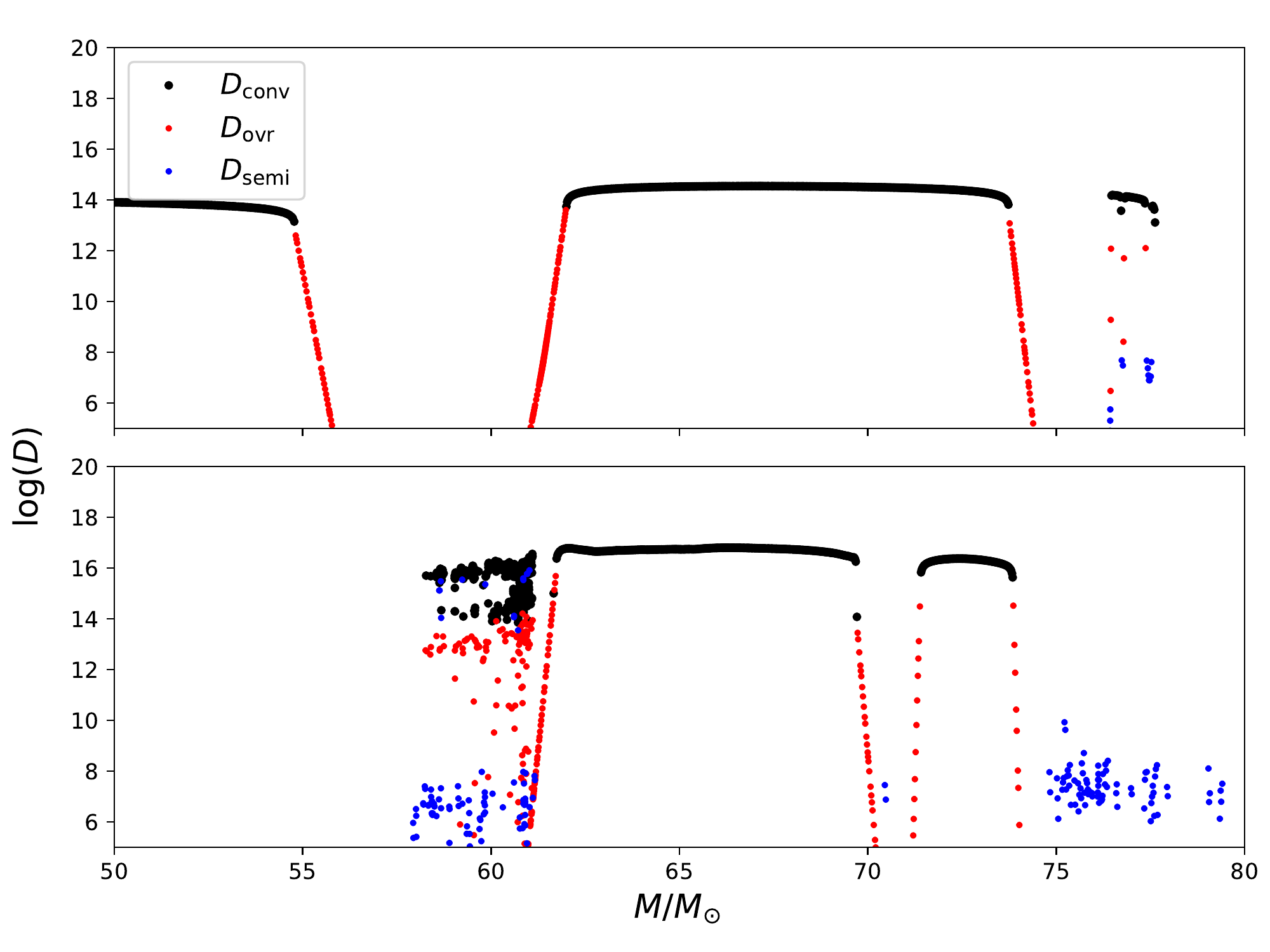}
    \caption{\textbf{Top}: Diffusion coefficients for for convection, semiconvection, and CBM in the \code{140Mledf-h} model $262\usp\unitstyle{yr}$ before the H-He interaction begins. \textbf{Bottom}: Same quantities at the time shown in \Fig{140Mother}. }
   \lFig{CRcoreD}
\end{figure}

\begin{figure}
	\includegraphics[width=\columnwidth]{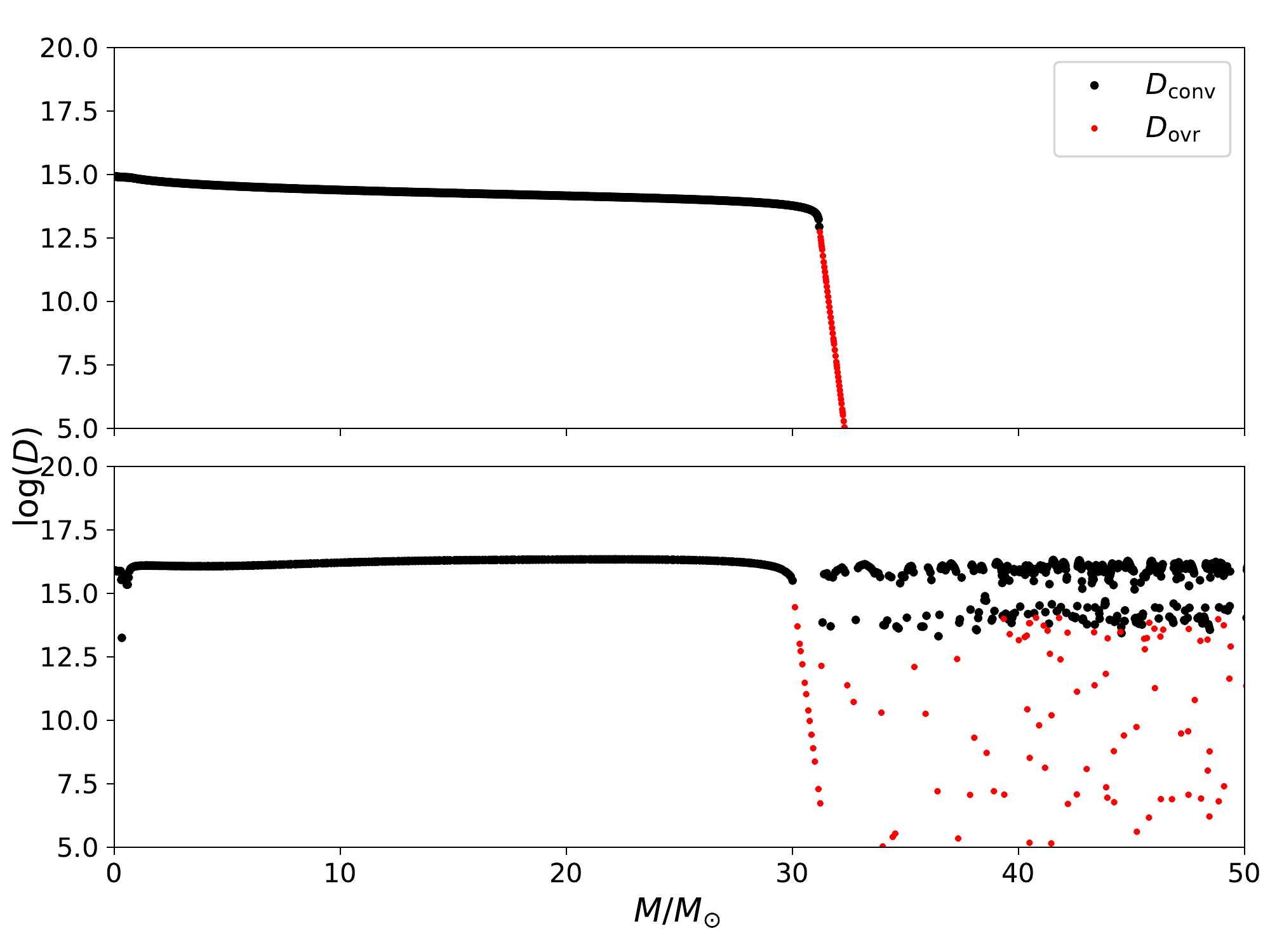}
    \caption{\textbf{Top}: Diffusion coefficients for convection and CBM in the \code{80Mschf-h} model $27\usp\unitstyle{yr}$ before the H-He interaction begins. \textbf{Bottom}: Same quantities at the time shown in \Fig{80mother}. }
   \lFig{CCcoreD}
\end{figure}


\bsp	
\label{lastpage}
\end{document}